\documentstyle[prd,eqsecnum,aps,twocolumn,epsfig]{revtex}

\catcode`\@=11

\def\maketitle2{\par 
\begingroup
\let\cite\@bylinecite
\def\thefootnote{\fnsymbol{footnote}}%
\twocolumn[\@maketitle2\vskip2pc]%
\thispagestyle{plain}\@thanks
\endgroup
\def\thefootnote{\arabic{footnote}}%
\setcounter{footnote}{0}%
\let\maketitle2\relax \let\@maketitle2\relax
\let\@thanks\relax \let\@authoraddress\relax \let\@title\relax
\let\@date\relax \let\thanks\relax \let\@abstract\relax 
\let\@pacs\relax}

\def\abstract#1{\gdef\@abstract{{\par 
\bgroup
\ifdim\prevdepth=-1000pt \prevdepth0pt\fi
\hsize\columnwidth
\dimen0=-\prevdepth \advance\dimen0 by17.5pt \nointerlineskip
\small\vrule width 0pt height\dimen0 \relax}{~~}#1\egroup}}

\def\pacs#1{\gdef\@pacs{{\par 
\bgroup
\hsize\columnwidth \parindent0pt
\ifdim\prevdepth=-1000pt \prevdepth0pt\fi
\dimen0=-\prevdepth \advance\dimen0 by20pt\nointerlineskip
\egroup} PACS numbers:~#1}}

\def\@maketitle2{
\@preprint
\@title
\ifdim\prevdepth=-1000pt \prevdepth0pt\fi
\@authoraddress
\@date
\begin{list}{}{\leftmargin=0.10753\textwidth \rightmargin=\leftmargin
\itemsep=1pc\partopsep=-1pc}
\item\@abstract
\item\@pacs
\end{list}
}

\catcode`\@=12

\newif\ifinlinefigures
\inlinefigurestrue
\newcommand{\bra}[1]{\langle #1 |}
\newcommand{\ket}[1]{| #1 \rangle}

\newcommand{\eq}[1]{eq.(\ref{#1})}
\newcommand{\dpar}[2]{\frac{\partial #1}{\partial #2}}
\newcommand{\vpar}[2]{\frac{\delta #1}{\delta #2}}

\newcommand{\bm}[1]{\mbox{\boldmath $#1$}}
\def\ltap{\raisebox{-.55ex}{\rlap{$\sim$}} \raisebox{.4ex}{$<$}}
\def\gtap{\raisebox{-.55ex}{\rlap{$\sim$}} \raisebox{.4ex}{$>$}}
\def\gsim{\mathrel{\gtap}}
\def\lsim{\mathrel{\ltap}}
\def\ghost#1{\vrule height#1 depth#1 width0pt \displaystyle}
\def\const{\mbox{const}}
\def\e{\mbox{e}}
\def\ch{\mbox{ch}}
\def\sh{\mbox{sh}}

\def\Re{\mathop{\mbox{Re}}}
\def\Im{\mathop{\mbox{Im}}}
\def\half{{1 \over 2}}

\begin{document}
\draft
\flushbottom

\title{False Vacuum Decay Induced by Particle Collisions}
\author{A.N.Kuznetsov and P.G.Tinyakov}
\address { Institute for Nuclear Research of the Russian
   Academy of Sciences,
   60th October Anniversary prospect,
   7a, Moscow 117312, Russia.}
\date{\today}
\abstract
{The semiclassical formalism for numerical calculation of the rate of
tunneling transitions induced by $N$ particles with total energy $E$
of order or higher than the height of the barrier is developed.  The
formalism is applied to the induced false vacuum decay in the massive
four-dimensional $-\lambda\phi^4$ model. The decay rate, as a function
of $E$ and $N$, is calculated numerically in the range $0.4\lsim
E/E_{sph}\lsim 3.5$ and $0.25 \lsim N/N_{sph}\lsim 1.0$, where
$E_{sph}$ and $N_{sph}$ are the energy and the number of particles in
the analog of the sphaleron configuration. The results imply that the
{\em two-particle} cross section of the false vacuum decay is
exponentially suppressed at least up to energies of order $10
E_{sph}$. At $E\sim E_{sph}$, this exponential suppression is
estimated as about 80\% of the zero energy suppression.}

\maketitle2
\narrowtext

\section{Introduction}

Non-perturbative effects related to tunneling play an important role
in many field theory models. Two widely known examples are the decay
of the metastable (false) vacuum in scalar models and winding number
transitions in sigma models and gauge theories at weak coupling.  At
low energies, these processes are well described by the semiclassical
approximation which relies upon the existence of classical (imaginary
time) solutions interpolating between initial and final states. In the
two examples mentioned above the interpolating solutions are
bounce~\cite{Bounce} and instanton~\cite{GaugeInstantons},
respectively. The Euclidean action of these solutions determines
the exponential part of the transition rate.
It is inversely proportional to the small coupling
constant; the transition rate is thus strongly suppressed.  In
realistic situations, the initial state may contain particles.  If
these particles have low energies, they can be taken into account
perturbatively and merely change the pre-exponential factor.

The situation is different if particles with parametrically high
energy are present in the initial state.  An example relevant to what
follows is high energy scattering in the false vacuum or in the
presence of the instanton. If the energy of the collision is of order
$1/\lambda$, where $\lambda$ is the small coupling constant, then the
application of the semiclassical approximation is not straightforward,
despite the fact that the two-particle initial state is
perturbative. The probability of induced tunneling transition may, and
in fact does depend exponentially on the energy of collision.
Quantitative description of this effect is the main purpose of this
paper.

The exponential dependence on collision energy is precisely what one
expects on physical grounds. The typical situation is illustrated in
Fig.\ref{decay-fig} which shows a generic profile of the potential
energy as a function of field for a model with false vacuum decay. The
false vacuum $\phi=0$ is separated from the true one by the potential
barrier whose height\footnote{Throughout this paper we somewhat
loosely use the term ``sphaleron'' not only for the particular static
solution in the Electroweak Theory~\cite{KLINKHAMER-MANTON}, but for
any static solution sitting in unstable equilibrium on top of the
potential barrier and thus characterizing its height. In the problem
of false vacuum decay the sphaleron is actually the critical
bubble~\cite{VOLOSHIN-KOBZAREV-OKUN}; $E_{sph}$ in the text stands for the
sphaleron energy. } $E_{sph}$ is parametrically $m/ \lambda$, where
$m$ is the mass scale in the model. The intuition based on quantum
mechanics suggests that at energies of order $E_{sph}$ the tunneling
suppression factor may be substantially reduced or even completely
absent.  This quantum mechanical intuition, however, should be
translated to field theory with care.

\ifinlinefigures
\begin{figure}[ht]
\begin{center}
\epsfig{file=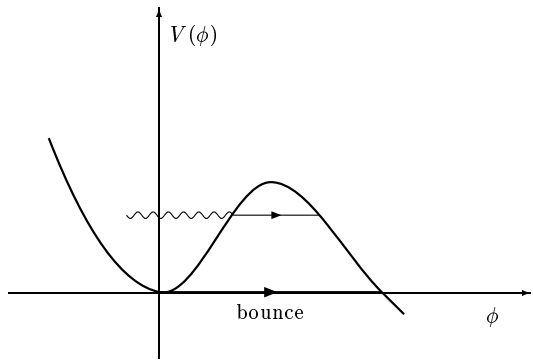,%
bbllx=60pt,bblly=620pt,%
bburx=310pt,bbury=800pt,%
width=210pt,height=150pt,%
clip=}
\end{center}
\caption{{\protect\small Generic picture of
false vacuum decay induced by particle collisions. The wavy line
represents an excited state above the false vacuum.} }
\label{decay-fig}
\end{figure}
\fi

The situation is much the same in the case of instanton transitions,
at least in models which possess a mass scale at the classical level.
An example is the Electroweak Theory where different topological
sectors are separated by the potential barrier of the height
$E_{sph}\sim 10 \mbox{TeV}$~\cite{KLINKHAMER-MANTON}. It was
found~\cite{RE} that tunneling between different sectors is enhanced in the
presence of high energy particles. Whether the exponential suppression
disappears completely at sufficiently high energy is still an open
question. Were it possible to observe such transitions in collider
experiments, the events would look quite spectacular due to
accompanying baryon and lepton number violation~\cite{t'Hooft} (for
recent review on baryon number violation see ref.\cite{RubShap}).

As was first noted in refs.\cite{RE}, at relatively low energies the
corrections to the tunneling rate can be calculated by perturbative
expansion in the background of the instanton (bounce). Further studies
showed that the actual expansion parameter is $E/E_{sph}$
\cite{McLVV,KRT2,Yaffe,ArnoldMattis} and the total cross section of
the induced tunneling has a remarkable form
\begin{equation}
\sigma_{tot} (E) \sim \exp \left\{
-{1\over\lambda} F_{HG}(E/E_{sph}) \right\} ,
\label{sigma=exp[F_HG]}
\end{equation}
where the function\footnote{The subscript HG stands for "holy grail"
\cite{reviews}.} $F_{HG}(\epsilon)$ is a
series in powers of $\epsilon$ (for a review see~\cite{reviews}).

While the perturbation theory in $\epsilon = E/E_{sph}$ is limited to
small $\epsilon$, the general form of the total cross section implies
that there might exist a semiclassical-type procedure which would
allow, at least in principle, to calculate $F_{HG}(\epsilon)$ at
$\epsilon\gsim 1$. Since the initial state of two highly energetic
particles is not semiclassical, the standard semiclassical procedure
does not apply and a proper generalization is needed. Such
generalization was proposed in refs.\cite{RT,T,RSTboundary}.  The
formalism of refs.\cite{RT,T,RSTboundary} reduces the calculation of
the exponential suppression factor to a certain classical boundary
value problem, whose analytical solution is not usually
possible\footnote{In some simple cases a non-trivial information about
the function $F_{HG}(E/E_{sph})$ can still be obtained by analytical
methods~\cite{SRLiuville}.}. The purpose of this paper is to develop
numerical techniques for calculating the function $F_{HG}(E/E_{sph})$
and to demonstrate how this techniques works in a simple model. As an
example we consider the induced false vacuum decay in the
$-\lambda\phi^4$ theory with the mass term (in a different context,
the problem of induced false vacuum decay was previously addressed in
refs.\cite{vac-decay,SRLiuville}). The rate of induced tunneling
transitions should be calculable by the same method in more realistic
models as well.

The motivation for this calculation is the following. Although general
arguments imply~\cite{unitarity} that the cross section of induced
tunneling is exponentially suppressed at all energies (our preliminary
numerical results~\cite{KT} agree with this statement), in realistic
models with finite coupling constant the possibility to observe
induced tunneling transitions depends on the {\em numerical} value of
the suppression factor. The only way to estimate this value is by
performing direct calculations. This may be of particular importance
in condensed matter models where coupling constants are often not very
small and even low energy transitions are observable.

The semiclassical approach proposed in refs.\cite{RT,T,RSTboundary} is
based on the conjecture that, with exponential accuracy, the
two-particle initial state can be substituted by a multiparticle one
provided that the number of particles is not parametrically large
(although not proven rigorously, this conjecture was checked
explicitly in several orders of perturbation theory in
$\epsilon$~\cite{Mueller}).
The few-particle initial state, in turn, can be
considered as a limiting case of truly multiparticle one with the
number of particles $N=\nu/\lambda$ when the parameter $\nu$ is sent
to zero. For the multiparticle initial state the transition rate is
explicitly semiclassical and has the form
\[
\sigma(E,N) \sim \exp \left\{
-{1\over\lambda} F(\epsilon,\nu) \right\} .
\]
According to the above conjecture, the function  $F_{HG}(\epsilon)$
can be reproduced in the limit $\nu\to 0$,
\[
\lim\limits_{\nu\to 0}F(\epsilon,\nu) = F_{HG}(\epsilon).
\]
Therefore, although indirectly, the function $F_{HG}(\epsilon)$ is
also calculable semiclassically.

Within the conventional semiclassical framework, the function
$F(\epsilon,\nu)$ is determined by the action evaluated on a
particular solution to the classical field equations with certain
boundary conditions~\cite{RSTboundary}. In this formulation, the
problem allows for numerical study. Namely, one can solve the
corresponding boundary value problem numerically and calculate the
function $F(\epsilon,\nu)$, which then can be used to extract
information about $F_{HG}(\epsilon)$. In the present paper we follow
this general strategy. On the lattice, it is not possible to actually
reach the point $\nu=0$, as the solution to the boundary value problem
becomes singular in this limit. It is important, however, that the
function $F(\epsilon,\nu)$ is monotonically decreasing function of
$\nu$ at fixed $\epsilon$~\cite{RT,T}, so that the point $\nu=0$
corresponds to maximum suppression at given $\epsilon$. If one finds
that $F(\epsilon,\nu)>0$ for some $\nu$, then
\[
F_{HG}(\epsilon) > F(\epsilon,\nu) > 0
\]
and the two-particle cross section is exponentially suppressed. The
energy $E_*$ as which the two-particle cross section would become
unsuppressed is the energy at which the line $F(\epsilon,\nu)=0$
would cross the $\nu=0$ axis.

In the $-\lambda\phi^4$ model with the mass term (described in detail
in Sect.3), we calculate the function $F(\epsilon,\nu)$ numerically in
the range $0.25<\nu<1$ and $0.4<\epsilon<3.5$, where $\nu$ and
$\epsilon$ are normalized to their values at the sphaleron.  We
perform extrapolation to $\nu=0$ and estimate the value of the
function $F_{HG}(\epsilon)$ at $\epsilon\sim 1$, i.e.\ at energies of
the order of the sphaleron energy. We find that at these energies the
zero energy exponential suppression of the two-particle cross section
is reduced by only about 20\%. By extrapolation of $F(\epsilon,\nu)$
to higher energies we set lower bound for the energy $E_*$ at which
the function $F_{HG}(\epsilon)$ may become zero and thus the
two-particle cross section may become unsuppressed. We find that
$E_*>10\,E_{sph}$.  Hence, the tunneling transitions induced by
particle collisions are exponentially suppressed well above the
sphaleron energy, at least in the model we consider.

The paper is organized as follows. In Sect.II we discuss in more
detail the definition of the multiparticle probability $\sigma(E,N)$
and the use of the semiclassical approximation for the calculation of
the corresponding exponent $F(\epsilon,\nu)$. We rewrite the formalism
of ref.\cite{RSTboundary} in a form suitable for numerical
calculations. In Sect.III the $-\lambda\phi^4$ model with the mass
term, as well as general features of the false vacuum decay in this
model, are described. We then turn in Sect.IV to analytical
calculation of the function $F(\epsilon,\nu)$ at $\epsilon \ll 1$ and
at those values of $\nu$ which maximize the transition rate. This is
done for comparison with numerical results. In Sect.V we present the
results of numerical calculation of the function
$F(\epsilon,\nu)$. Sect.VI contains discussion and concluding
remarks. The details of numerical techniques are given in Appendices A
and B.

\section{Semiclassical calculation of inclusive multiparticle probability}

In this Section we review the the formalism of
refs.\cite{RT,T,RSTboundary} and rewrite it in the form suitable for
numerical calculations.

The inclusive multiparticle probability $\sigma(E,N)$ discussed
in the Introduction is defined as follows,
\begin{equation}
\sigma(E,N) = \sum_{i,f} |\bra{f} \hat{S} \hat{P}_E \hat{P}_N \ket{i}|^2,
\label{sigmaEN}
\end{equation}
where $\hat{S}$ is the S-matrix, $\hat{P}_{E,N}$ are projectors onto
subspaces of fixed energy $E$ and fixed number of particles $N$,
respectively, while the states $\ket{i}$ and $\ket{f}$ are
perturbative excitations above two vacua lying on different sides of
the barrier. $\sigma(E,N)$ can be interpreted as the total probability
of tunneling from a state of energy $E$ and number of particles $N$,
summed over all such states with equal weight. 

The advantage of dealing with $\sigma(E,N)$ instead of
$\sigma_{tot}(E)$ is that the r.h.s. of \eq{sigmaEN} can be written in
the functional integral form, in which the semiclassical approximation
is equivalent to the saddle-point integration. The double path
integral representation for $\sigma(E,N)$ reads~\cite{RT}
\begin{eqnarray}
\sigma(E,N) &=& \int d\theta dT da_{\bf k} da^*_{\bf k}
db_{\bf k} db^*_{\bf k} d\phi(x) d\phi'(x)
\exp \Bigl\{ -iN\theta
\nonumber\\
&&-iET
-\int d\bm{k} a_{\bf k} a^*_{\bf k} \e^{-i\theta - i\omega_kT}
-\int d\bm{k} b_{\bf k} b^*_{\bf k}
\nonumber\\
&&+ B_i(a_{\bf k},\phi_i) + B_f(b^*_{\bf k},\phi_f)
+ B^*_i(a^*_{-\bf k},\phi_i') 
\nonumber\\
&&+ B^*_f(b_{-\bf k},\phi_f')
+ i S(\phi) -i S(\phi') \Bigr\} ,
\label{PIsigmaE,N}
\end{eqnarray}
where the boundary terms $B_i$ and $B_f$ are
\begin{eqnarray}
B_i(a_{\bf k},\phi_i) &=&\half  \int\! d\bm{k} \biggl[
- \omega_{\bf k} \phi_i(\bm{k})\phi_i(-\bm{k})
-  a_{\bf k} a_{\bf -k} e^{-2i\omega_kT_i}
\nonumber\\&&
+ 2\sqrt{2\omega_{\bf k}}\, e^{-i\omega_kT_i}
a_{\bf k} \phi_i(\bm{k}) \biggr] ,
\nonumber\\
B_f(b^*_{\bf k},\phi_f) &= & \half \int\! d\bm{k}\biggl[
-  \omega_{\bf k}
\phi_f(\bm{k}) \phi_f(-\bm{k})
-  b^*_{\bf k} b^*_{\bf -k}e^{2i\omega_kT_f}
\nonumber\\&&
+  2\sqrt{2\omega_{\bf k}}e^{i\omega_kT_f}
b^*_{\bf k} \phi_f(-\bm{k}) \biggr] .
\label{Bi,f}
\end{eqnarray}
Here $\phi_{i,f}(\bm{k})$ are the spatial Fourier transforms of the
field at initial and final times $T_i$ and $T_f$, respectively.  The
limit $T_{i,f}\to\mp\infty$ is assumed at the end of the calculation.
The complex integration variables $a_{\bf k}$ and $b^*_{\bf k}$ come
from coherent state representation of initial and final
states~\cite{FadSlav};
they are classical counterparts of annihilation and
creation operators.  The integration over these variables substitutes
the summation over initial and final states in \eq{sigmaEN}. The
functional integrals over $\phi(x)$ and $\phi'(x)$ come from the
amplitude and complex conjugate amplitude, respectively. The
integrations include the boundary values $\phi_{i,f}$ and
$\phi'_{i,f}$.

From \eq{PIsigmaE,N} it follows immediately that in the weak coupling
limit $\sigma(E,N)$ has the semiclassical form (\ref{sigmaEN}). Indeed,
changing the integration variables
\[
\{ a,b,\phi,\phi'\} \to {1\over\sqrt{\lambda}} \{a,b,\phi,\phi'\}
\]
and taking into account that after this transformation the action
becomes proportional to $1/\lambda$, we arrive at \eq{sigmaEN} where
$\epsilon = \lambda E$, $\nu=\lambda N$ and the function
$F(\epsilon,\nu)$ is determined by the saddle-point value of the
exponent in \eq{PIsigmaE,N}.

Let us now discuss the saddle-point equations for the integral
(\ref{PIsigmaE,N}). We will see that these equations reduce to certain
boundary value problem for the fields $\phi$ and $\phi'$.  The
variables $a_{\bf k}$, $a^*_{\bf k}$, $b_{\bf k}$ and $b^*_{\bf k}$ do
not play any role in what follows. Moreover, they enter the exponent
quadratically. Integrating them out we find
\begin{eqnarray}
\sigma(E,N) &=& \int d\theta dT d\phi(x) d\phi'(x)
\prod\limits_{\bf k} \delta (\phi_f(\bm{k}) - \phi'_f(\bm{k}))
\nonumber\\&&
\times \exp \biggl\{ -iN\theta -iET + i S(\phi) -i S(\phi')
\nonumber\\&&
- \half \int d\bm{k} { \omega_{\bf k} \over 1 - \gamma_{\bf k}^2 }
\Bigl( (1+\gamma_{\bf k}^2)
 [\phi_i(\bm{k}) \phi_i(-\bm{k}) 
\nonumber\\&&
+ \phi'_i(\bm{k}) \phi'_i(-\bm{k})]
  - 4 \gamma_{\bf k} \phi_i(\bm{k}) \phi'_i(-\bm{k})
\Bigr)
\biggl\} ,
\label{smallPIsigmaEN}
\end{eqnarray}
where
\[
\gamma_{\bf k} = \e^{i\theta+i\omega_k T}.
\]
The important feature of the representation (\ref{smallPIsigmaEN}) is
that the exponent in the r.h.s. contains only the action and the
boundary values of the fields. Thus, the discretization of this
expression is straightforward.

Let us turn to the saddle point equations. Varying the exponent with
respect to the fields $\phi(x)$ and $\phi'(x)$ we find
\begin{equation}
\vpar{S}{\phi} = \vpar{S}{\phi'} =0,
\label{field-eqs}
\end{equation}
i.e.\ the usual field equations. The boundary conditions for these
equations come from the variation with respect to the boundary values
of the fields. At $t=T_f$, because of the $\delta$--function, the
variations are subject to the constraint $\delta\phi_f(\bm{x}) =
\delta\phi'_f(\bm{x})$.  Since $\delta S /\delta \phi(T_f,\bm{x}) =
\dot\phi(T_f,\bm{x})$ we obtain
\begin{eqnarray}
\dot\phi(T_f,\bm{x}) & = & \dot\phi'(T_f,\bm{x}),\nonumber \\
\phi(T_f,\bm{x}) & = & \phi'(T_f,\bm{x}).
\label{bc-1}
\end{eqnarray}
Thus, in the final asymptotic region the saddle-point fields $\phi$
and $\phi'$ coincide. Note that the dependence on $T_f$ cancels out in
the difference $S(\phi)-S(\phi')$.

The variation with respect to $\phi_i$ and $\phi'_i$ leads to two
equations which can be written in the following form,
\begin{eqnarray}
i\dot\phi_i(\bm{k}) + \omega_{\bf k} \phi_i(\bm{k}) & = &
\gamma_{\bf k} \left( i\dot\phi'_i(\bm{k})
+ \omega_{\bf k} \phi'_i(\bm{k}) \right), \nonumber\\
-i\dot\phi_i(\bm{k}) + \omega_{\bf k} \phi_i(\bm{k}) & = &  
{1\over \gamma_{\bf k} } \left( -i\dot\phi'_i(\bm{k}) + \omega_{\bf k}
\phi'_i(\bm{k}) \right).
\label{bc-2}
\end{eqnarray}
Let us check that these boundary conditions imply the independence of
the exponent in \eq{smallPIsigmaEN} of the initial normalization time
$T_i$.  The action depends on $T_i$ explicitly, while the boundary
term in \eq{smallPIsigmaEN} depends on $T_i$ through the boundary
values of the fields, $\phi_i$ and $\phi'_i$. Assuming that the fields
linearize and satisfy free field equation in the initial asymptotic
region, and integrating by parts in the action, the $T_i$--dependent
contributions in the exponent in \eq{smallPIsigmaEN} read
\begin{eqnarray}
- \half \int d\bm{k} \biggl[ { \omega_{\bf k} \over 1 - \gamma_{\bf k}^2 }
\bigl[ (1+\gamma_{\bf k}^2)
 (\phi_i(\bm{k}) \phi_i(-\bm{k}) + \phi'_i(\bm{k})
\phi'_i(-\bm{k})) \nonumber \\
  - 4 \gamma_{\bf k} \phi_i(\bm{k}) \phi'_i(-\bm{k})
\bigr]
+i\dot\phi_i(\bm{k})\phi_i(-\bm{k}) - i\dot\phi'_i(\bm{k}) \phi'_i(-\bm{k})
\biggr].\nonumber
\end{eqnarray}
By making use of eqs.(\ref{bc-2}) it is straightforward to see
that the integrand in this expression vanishes.

Finally, there are two saddle-point equations which come from the
variation of the exponent in \eq{smallPIsigmaEN} with respect to $\theta$
and $T$. They read
\begin{eqnarray}
E &=& \int {2\omega_{\bf k}^2 \gamma_{\bf k} d\bm{k} 
	\over (1-\gamma_{\bf k}^2)^2}
[\phi_i(\bm{k})-\gamma_{\bf k}\phi'_i(\bm{k})]
[\phi'_i(-\bm{k})-\gamma_{\bf k}\phi_i(-\bm{k})],
\nonumber \\
N &=& \int  {2\omega_{\bf k} \gamma_{\bf k} d\bm{k}
	\over (1-\gamma_{\bf k}^2)^2}
[\phi_i(\bm{k})-\gamma_{\bf k}\phi'_i(\bm{k})]
[\phi'_i(-\bm{k})-\gamma_{\bf k}\phi_i(-\bm{k})].
\nonumber \\ ~&~&~
\label{ENspec}
\end{eqnarray}
These equations determine the saddle-point values of $\theta$ and $T$ in
terms of $E$ and $N$.

The initial boundary conditions (\ref{bc-2}) simplify when written in
terms of frequency components. In the initial asymptotic region where
$\phi$ and $\phi'$ are free fields, we can write
\begin{eqnarray}
\phi(x) &=& \int { d{\bm k} \over \sqrt{(2\pi)^3 2\omega_k} }
\Bigl\{ f_{\bf k} \e^{-i\omega_kt+i{\bf kx}}
+ g_{\bf k}^* \e^{i\omega_kt-i{\bf kx}}  \Bigr\} ,
\nonumber\\
\phi'(x) &=& \int { d{\bm k} \over \sqrt{(2\pi)^3 2\omega_k} }
\Bigl\{ f'_{\bf k} \e^{-i\omega_kt+i{\bf kx}}
+ {g'}_{\bf k}^* \e^{i\omega_kt-i{\bf kx}}  \Bigr\} .
\nonumber\\ & &
\label{asymptotic form}
\end{eqnarray}
Eqs.(\ref{bc-2}) then become
\begin{eqnarray}
f_{\bf k} &=& \gamma_{\bf k} f'_{\bf k},
\nonumber\\
g^*_{\bf k} &=& {1\over \gamma_{\bf k}} {g'}^*_{\bf k},
\label{simplebc-2}
\end{eqnarray}
while the saddle-point equations (\ref{ENspec}) read
\begin{eqnarray}
E &=& \int d\bm{k} \omega_{\bf k} f_{\bf k} g^*_{\bf k},
\nonumber\\
N &=& \int d\bm{k} f_{\bf k} g^*_{\bf k}.
\label{ENspec2}
\end{eqnarray}
One may recognize the usual expressions for the energy and the number
of particles contained in the free classical field, $f_{\bf k}
g^*_{\bf k}$ being the occupation number in the mode with
spatial momentum $\bm{k}$.

Since, as follows from \eq{bc-1}, the fields $\phi$ and $\phi'$
coincide in the final asymptotic region and thus coincide everywhere,
there are solutions to the boundary value problem (\ref{field-eqs}) -
(\ref{bc-2}) which have $\gamma_{\bf k}=1$ and $\phi=\phi'$ in the
initial asymptotic region. These solutions do not describe tunneling
transitions --- for them the exponent in \eq{smallPIsigmaEN} vanishes.
They correspond to classical propagation over the barrier (recall that
we require that the initial and final states are on different sides of
the barrier) and clearly exist only at $E>E_{sph}$.  If such
over-barrier solutions exist for some values of $E$ and $N$, then the
corresponding multiparticle probability is definitely not
exponentially suppressed. Conversely, if the two-particle cross
section are not exponentially suppressed at some energy, then at this
energy there should exist over-barrier solutions with arbitrarily
small $N$. This observation provides a strategy of search for
unsuppressed induced transitions~\cite{Rebbi-Singleton}. Our formalism
is not suitable for studying these solutions as the boundary
conditions \eq{bc-1} and \eq{bc-2} are degenerate in the case
$\phi_i=\phi'_i$, $\gamma_{\bf k} =1$.

The solution to the boundary value problem
(\ref{field-eqs})-(\ref{bc-2}) which describes {\em tunneling}
transition is obtained when, in the initial asymptotic region, $\phi$
and $\phi'$ are treated as values of the same field $\phi$ taken on
{\em different sheets} in the complex time plane. The existence of
such solutions is suggested by the analogy with barrier penetration in
one-dimensional quantum mechanics (below we show by numerical
calculations that analogous solutions exist in the field theory as
well, and argue that their interpretation as tunneling solutions is
consistent).

\ifinlinefigures
\begin{figure}[ht]
\begin{center}
\epsfig{file=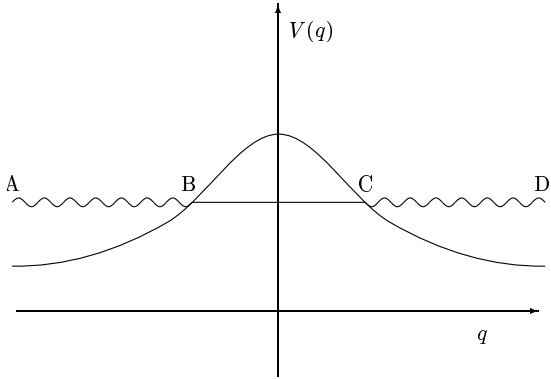,%
bbllx=60pt,bblly=620pt,%
bburx=310pt,bbury=800pt,%
width=210pt,height=150pt,%
clip=}
\end{center}
\caption{{\protect\small Barrier penetration in one-dimensional
quantum mechanics. The classical solution with given energy, considered
in the complex time plane, describes both evolution in the classically
forbidden (BC part) and classically allowed regions (AB and CD parts).}
}
\label{QMtun-fig}
\end{figure}
\fi
To illustrate this idea consider barrier penetration in one-dimensional
quantum mechanics with the potential
\[
V(q) = {1\over\ch^2q}
\]
shown in Fig.\ref{QMtun-fig}. At energy $E<1$ there exist two
classically allowed regions, AB and CD, separated by the classically
forbidden one, BC. The classical solution with energy $E$ is given by
the following equation,
\begin{equation}
\sqrt{{E\over 1-E}} \sh q = \ch (\sqrt{2E} t).
\label{QMsolution}
\end{equation}
As defined by this equation, $q(t)$ is an analytic (except for
isolated points, see below) function which is periodic in Euclidean
time with the period $T=2\pi/\sqrt{2E}$ and real at purely imaginary
time $\Re t=0$ where it describes oscillations between the two turning
points B and C.  It is also real on the lines $\Im t = nT/2$, $n \in
Z$ where it represents the motion in the classically allowed regions:
in the region CD for even $n$ and in the region AB for odd $n$. The
entire evolution from $q=-\infty$ to $q=+\infty$, including tunneling,
corresponds to the contour ABCD in the complex time plane as shown in
Fig.\ref{contour-fig}. The solution (\ref{QMsolution}) taken on the
contour ABCD contributes to the amplitude, while taken on the complex
conjugate contour A$'$B$'$CD it contributes to the complex conjugate
amplitude. These two parts of the solution are quantum-mechanical
analogs of the fields $\phi$ and $\phi'$. In perfect analogy to
\eq{smallPIsigmaEN}, the transition probability can be written as
$\exp\{ iS_{ABCD} - i S_{A'B'CD} + ET\}$, where subscripts indicate
the contour along which the action is evaluated. In the sum of the
actions the Minkowskian parts of the contours cancel, while the
Euclidean parts add and give $iS_{BB'} = - S_E(T)$, where $S_E(T)$ is
the Euclidean action per period. Thus, we recover the standard WKB
transition probability $\exp\{ ET - S_E(T)\}$.

\ifinlinefigures
\begin{figure}[ht]
\begin{center}
\epsfig{file=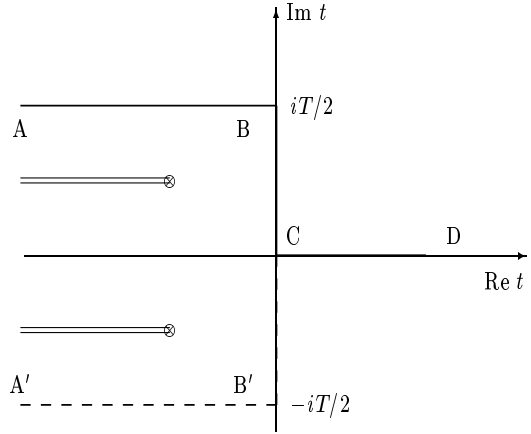,%
bbllx=50pt,bblly=620pt,%
bburx=310pt,bbury=800pt,%
width=210pt,height=170pt,%
clip=}
\end{center}
\caption{{\protect\small The contour in the complex time plane where
the boundary value problem (6)--(8) is
formulated. Crossed circles represent singularities of the field.}}
\label{contour-fig}
\end{figure}
\fi
For this picture to be consistent, there must be singularities in the
complex time plane which prevent the deformation of the contour ABCD to
the real time axis. In particular, the initial asymptotic region A should
be separated from the real time axis by a cut. It is straightforward to
check that the solution (\ref{QMsolution}) indeed has the right structure
of singularities. The function $q(t)$ itself is finite, but the derivative
\[
{dq\over dt} = {\sqrt{2E} \sh \left( \sqrt{2E} t \right) \over
\sqrt{ {E\over 1-E} + \ch^2\left( \sqrt{2E} t\right) } }
\]
is infinite at the points $t_*$ satisfying
\begin{equation}
{E\over 1-E} + \ch^2\left( \sqrt{2E} t_*\right) =0,
\label{t*equation}
\end{equation}
i.e.\ at the points
\begin{equation}
t_* = \pm {1\over 2\sqrt{2E}} \mbox{arccosh} {1+E\over 1-E} +
i {\pi\over \sqrt{2E}} (n+1/2); ~~~~n \in Z.
\label{t*=}
\end{equation}
Note that, as follows from eqs.(\ref{QMsolution}), (\ref{t*equation}),
the corresponding values $q_*$ satisfy
\[
\ch q_* =0.
\]
The type of the singularity can be found by solving \eq{QMsolution} in
the vicinity of the singular point, $t=t_* + \delta t$, $q=q_* + \delta
q$. Expanding \eq{QMsolution} to the leading order we find
\[
(\delta q)^2 = \pm 2\sqrt{2} \delta t.
\]
Thus, the singularities are the square root branching points. The cuts
relevant for our boundary value problem are shown in
Fig.\ref{contour-fig}.

In the spirit of this quantum mechanical example, we interpret the fields
$\phi$ and $\phi'$ in the initial asymptotic region as values of the same
field $\phi$ taken on different sheets in the complex time plane. The
fields $\phi$ and $\phi'$ are obtained by analytical continuation (in the
initial asymptotic region) to the real time axis from the two complex
conjugate contours ABCD and A$'$B$'$CD. It is convenient to reformulate
the boundary value problem directly in terms of the fields on these
contours (note that the analytical continuation in the initial asymptotic
region can be done explicitly by means of eqs.(\ref{asymptotic form})). Let
us assume for a moment that the saddle point values of $T$ and $\theta$
are purely imaginary and positive. Then
\begin{equation}
\gamma_{\bf k} = \e^{-\theta - \omega_kT},
\label{gamma}
\end{equation}
and the analytical continuation in eqs.(\ref{bc-2})--(\ref{ENspec2})
from the real time axis, where they are originally formulated, to the
contours ABCD and A$'$B$'$CD results in the substitution of $\gamma_{\bf
k}$ by
\[
\gamma = \e^{-\theta}.
\]
In other words, eqs.(\ref{bc-2})--(\ref{ENspec2}) remain valid if
simultaneously with the analytical continuation the substitution
$\gamma_{\bf k} \to \gamma$ is performed. Note that in this
formulation the parameter $T$ enters the boundary conditions
implicitly as the total amount of Euclidean evolution. It is worth
noting also that, except for the asymptotic region, the contours ABC
and A$'$B$'$C can be arbitrarily deformed, provided the singularities
are not crossed.

In general, there may be several saddle points in the integral
(\ref{smallPIsigmaEN}). In this case, the physically relevant one is
that continuously connected to the saddle point which emerges in the
low energy perturbation theory~\cite{T}. The perturbative saddle point
has purely imaginary $\theta$ and $T$. Moreover, it obeys the property
\begin{equation}
[\phi(t,\bm{x})]^* =\phi(t^*,\bm{x}).
\label{conjugation-eq}
\end{equation}
In particular, $\phi$ is real on the real time axis. Unless a
bifurcation occurs, these properties should hold for the
non-perturbative saddle point as well. In numerical calculations we
have not found any bifurcation points\footnote{The numerical algorithm
we use (see Sect.5) stops to converge in the vicinity of bifurcation
points.}, so in what follows we assume that these are absent
and consider saddle points obeying \eq{conjugation-eq}.

Eq.(\ref{conjugation-eq}) implies that on the contour A$'$B$'$CD the
field is complex conjugate of that on the contour ABCD. Thus,
the field is real on the part CD of the contour, and only the
part ABC of the contour needs to be considered. In terms of the field
on the contour ABC, the multiparticle probability reads
\[
\sigma(E,N) \sim
\exp \left\{ - {1\over\lambda} F(\epsilon,\nu) \right\},
\]
\begin{eqnarray}
- {1\over\lambda} F(\epsilon,\nu) &=&
N\theta + ET + 2\Re [i S_{ABC}(\phi)]
\nonumber\\
& & 
- {1-\gamma \over 1+\gamma}  \int d\bm{k} \omega_{\bf k}
\Re\phi_i(\bm{k}) \Re\phi_i(-\bm{k})
\label{saddle-sigmaEN}
\\
& & 
+ {1+\gamma \over 1-\gamma} \int d\bm{k} \omega_{\bf k}
\Im\phi_i(\bm{k}) \Im\phi_i(-\bm{k}) ,
\nonumber
\end{eqnarray}
where $\gamma$ is given by \eq{gamma}. The field $\phi$ satisfies the
following boundary value problem formulated on ABC,
\begin{mathletters}
  \begin{eqnarray}
  \label{final_BVP}
  \vpar{S}{\phi} &=& 0,
  \label{BVP-fieldeq} \\
  \Im \dot \phi(0,\bm{x}) &=& \Im \phi(0,\bm{x}) = 0 ,
  \label{BVP-bc1} \\
  f_{\bf k} &=& \e^{-\theta} g_{\bf k},
  \label{BVP-bc2}
  \end{eqnarray}
\end{mathletters}
where $f_{\bf k}$ and $g_{\bf k}$ are frequency components of the
field in the asymptotic region A along AB. Note that \eq{BVP-bc1}
implies reality of the field on the real time axis. It consists of two
real equations. Eq.(\ref{BVP-bc2}) also consists of two real
conditions imposed on the field and its derivative in the asymptotic
region A. In total, there are two real (one complex) boundary
conditions at each end of the contour, so the boundary value problem
is completely specified. Finally, eqs.(\ref{ENspec2}) which determine
the two auxiliary parameters $T$ and $\theta$, become
\begin{eqnarray}
N &=& \int d\bm{k} \omega_{\bf k} \biggl(
{2 \gamma \over (1+\gamma)^2} \Re \phi_i(\bm{k}) \Re \phi_i(-\bm{k}) 
\nonumber\\
& &+
{2 \gamma \over (1-\gamma)^2} \Im \phi_i(\bm{k}) \Im \phi_i(-\bm{k}) 
\biggr), \nonumber\\
E &=& \int d\bm{k} \omega_{\bf k}^2 \biggl(
{2 \gamma \over (1+\gamma)^2} \Re \phi_i(\bm{k}) \Re \phi_i(-\bm{k})
\nonumber\\
& & +
{2 \gamma \over (1-\gamma)^2} \Im \phi_i(\bm{k}) \Im \phi_i(-\bm{k}) 
\biggr).
\label{finalENspec}
\end{eqnarray}
This is the boundary value problem we solve numerically in the present
paper.

The interpretation of solutions to the boundary value problem
(\ref{final_BVP}) is the following. On the part CD of the contour, the
saddle-point field is real; it describes the evolution of the system
after tunneling. On the contrary, it follows from boundary conditions
(\ref{BVP-bc2}) that in the initial asymptotic region the saddle-point
field is complex provided that $\theta\neq 0$. Thus, the initial state
which maximizes the probability (\ref{PIsigmaE,N}) is not described by
a classical field, i.e.\ this stage of the evolution is essentially
quantum even at $N\sim 1/\lambda$.

The case $\theta=0$ is exceptional. In this case, the boundary
condition (\ref{BVP-bc2}) reduce to the reality conditions imposed at
$\Im t =T/2$. The solution to the resulting boundary value problem is
given by periodic instanton of ref.\cite{KRTperiod}. Periodic
instanton is a real periodic solution to the Euclidean field equations
with period $T$ and two turning points at $t=0$ and $t=iT/2$. Being
analytically continued to the Minkowskian domain through the turning
points, periodic instanton is real at the lines $\Im t =0$ and $\Im t
= T/2$ and therefore satisfies the boundary value problem
(\ref{final_BVP}) with $\theta=0$. Periodic instanton is the close
analog of the quantum-mechanical solution discussed above.

\section{The model}

We perform numerical calculations in the model of one real scalar
field $\phi$, defined by the Minkowskian action
\begin{equation}
S = \int d^4 x \Bigl( \half \partial_{\mu} \phi\partial^{\mu} \phi
- \half m^2\phi^2   + {1\over 4}\lambda\phi^4 \Bigr),
\label{model_action}
\end{equation}
where $\lambda$ is a {\em positive} constant. In the case $m=0$ and
infinite volume, the state $\phi=0$ is stable classically
but meta-stable with respect to quantum fluctuations. The decay of this
state is described by the well-known instanton
solution~\cite{Fubini,Lipatov},
\begin{equation}
\phi_{inst} = 2\sqrt{{2\over \lambda}} {\rho\over x^2+\rho^2} .
\label{instanton}
\end{equation}
Due to conformal invariance, the instanton can have arbitrary size
$\rho$. The instanton action,
\begin{equation}
S_{inst} = {8 \pi^2 \over 3\lambda},
\label{Sinst}
\end{equation}
is independent of $\rho$. At the same time, the energy barrier between
the state $\phi=0$ and the instability region depends on the
configuration size and tends to zero as $\rho\to\infty$.

At $m\neq 0$, the conformal invariance is softly broken and regular
Euclidean solutions with finite action are forbidden by scaling
argument. The action of the instanton-like configuration depends on
its size and is minimal at $\rho=0$. The decay of the state $\phi=0$
is dominated by a set of approximate solutions, constrained
instantons~\cite{Affleck},
minimizing the Euclidean action under the constraint
that their size is $\rho$. At $\rho^2 m^2\ll 1$, the field of the
constrained instanton coincides with the massless solution
(\ref{instanton}) at $x\ll m^{-1}$ and falls off exponentially at
$x\gsim m^{-1}$. At low energies, contribution of small size
constrained instantons is dominant and the probability of the decay is
determined by the factor $\exp(-S_{inst})$.

At $m\neq 0$, the barrier separating the state $\phi=0$ from
instability region $\phi\gsim m/\sqrt{\lambda}$ is finite.
There exists the sphaleron solution characterizing the barrier height.
It is a static O(3)--symmetric configuration $\phi_{sph}(r)$
satisfying the equation
\[
-\phi''-{2\over r} \phi' +m\phi -\lambda\phi^3 =0,
\]
which can be solved numerically by shooting method (see,
e.g., ref.\cite{NumRec}). The energy of the sphaleron equals
\[
E_{sph}  =  \ghost{14pt} 18.9 {m\over\lambda}.
\]
Since the sphaleron corresponds to the top of the potential barrier, it
is unstable and has one negative mode. In the model
(\ref{model_action}) the negative eigenvalue can be found numerically,
\[
\omega^2_- = - 15.3 m^2.
\]
Being slightly perturbed in the direction of the negative mode, the
sphaleron rolls down to the metastable vacuum and decays into particles
(i.e., becomes a collection of plane waves). In this way the number of
particles contained in the sphaleron is
defined~\cite{sphaleron-decay}.
In numerical simulations it can be measured
by making use of eqs.(\ref{asymptotic form}),(\ref{ENspec2}).
In the model (\ref{model_action}) we found 
\[
N_{sph}  =  \ghost{14pt} 10.5 {1\over\lambda}.
\]

The peculiar feature of the model (\ref{model_action}) is the absence
of a stable vacuum state. However, this is not essential in
semiclassical calculations. One may imagine that the term $\alpha
\phi^6$ is added to the potential with a small coefficient $\alpha$.
Such term would create a true vacuum at $\phi\sim
\sqrt{\lambda/\alpha}$ but would not change equations of motion in the
region $\phi\leq m/\sqrt{\lambda}$. In numerical calculations it is
more convenient not to add such a term since the singularity in the
final state (i.e.\ on the part CD of the contour of
Fig.\ref{contour-fig}) is a clear sign of the false vacuum decay.

Before turning to the calculation of the function $F(\epsilon,\nu)$ in
this model, it is useful to note that neither of the two parameters
entering the action (\ref{model_action}) appears in the classical
equations of motion. Indeed, after changing the variables according to
\begin{eqnarray}
x &\to& m^{-1}x,
\nonumber\\
\phi &\to& {m\over \sqrt{\lambda}}\phi,
\nonumber
\end{eqnarray}
the action (\ref{model_action}) takes the following form,
\[
S = {1\over \lambda} \int d^4 x \Bigl( \half \partial_{\mu}
\phi\partial^{\mu} \phi - \half \phi^2 + {1\over 4}\phi^4 \Bigr)
\]
It is also clear from this equation that the coupling constant
$\lambda$ controls the semiclassical approximation.

\section{Perturbation theory at $\epsilon\ll 1$ and $\theta=0$}

At low energies the function $F(\epsilon,\nu)$ can be calculated
analytically. For the purpose of comparison with numerical results, we
perform this calculation in the case $\theta=0$, i.e.\ for the periodic
instanton.

\ifinlinefigures
\begin{figure}[ht]
\begin{center}
\epsfig{file=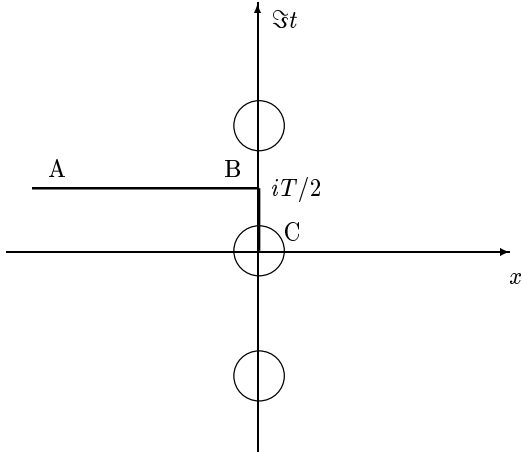,%
bbllx=60pt,bblly=610pt,%
bburx=280pt,bbury=800pt,%
width=210pt,height=180pt,%
clip=}
\end{center}
\caption{{\protect\small Low energy periodic instanton is a sum of
constrained instantons sitting at the distance $T$ along the imaginary
time axis.} }
\label{Pinst-fig}
\end{figure}
\fi
At $\epsilon\ll 1$, the periodic instanton can be approximated by an
infinite chain of zero energy constrained instantons of the size
$\rho$ sitting at the distance $T$ along the imaginary time axis, as
shown in Fig.\ref{Pinst-fig}. The action per period for this field
configuration, $S_p$, is a function of $\rho$ and $T$. Since, by
reflection symmetry, the point $t=iT/2$ is a turning point,
$\dot\phi(iT/2,\bm{x})=0$, the field is real on the line AB of the
contour of Fig.\ref{Pinst-fig}. Thus, it satisfies the boundary
condition \eq{BVP-bc2} with $\theta=0$. One also has
\[
2 \Re[iS_{ABC}(\phi)]=-S_p(\rho,T),
\]
while the boundary term in the exponent in \eq{saddle-sigmaEN}
vanishes. Therefore, for the case of periodic instanton, the exponent
in \eq{saddle-sigmaEN} reads
\begin{equation}
-{1\over\lambda} F(\epsilon,\nu_p(\epsilon)) =
ET - S_p(\rho, T),
\label{Fperinst}
\end{equation}
where the number of particles in the initial state, $\nu_p(\epsilon)$,
is the function of energy determined by the first of
eqs.(\ref{finalENspec}).

For this field configuration to be an (approximate) solution to the
equations of motion, the exponent must be stationary with respect to
$\rho$ and $T$,
\begin{eqnarray}
\dpar{S_p}{\rho} &=& 0,
\nonumber\\
\dpar{S_p}{T} &=& E.
\label{stationarityS_p}
\end{eqnarray}
These equations determine the values of $\rho$ and $T$ in terms of
energy.

The field of the periodic instanton is the sum of contributions of
individual constrained instantons. The field of a single constrained
instanton located at the point $\tau=\tau_0$, where $\tau=\Im t$ is
the Euclidean time, can be approximated as follows (throughout this
section we use units $m=1$),
\[
\phi_0(\tau-\tau_0,{\bm x}) = \sqrt{8 \over \lambda } \rho {K_1(R)\over R}.
\]
Here $K_1$ is the modified Bessel function and $R=\sqrt{(\tau-\tau_0)^2
+ {\bm x}^2+\rho^2}$. Hence, the periodic instanton field is
\begin{eqnarray}
\phi_p(x) &=& \sum\limits_{n} \phi_0 (\tau-nT, {\bm x})
\nonumber\\
&=&  \phi_0 (\tau, {\bm x}) + \sqrt{8\over\lambda}{\rho \over r}
\int\limits_0^{\infty}  k\,dk 
{ \sin(kr) \ch(\omega_{\bf k}\tau)
\over\omega_{\bf k} \,(\e^{\omega_kT}-1) },
\end{eqnarray}
where $r=\sqrt{\bm{x}^2}$ and $|\tau|\leq T/2$. Substituting this
expression into the Euclidean version of the action
(\ref{model_action}) and evaluating the time integral over the period
one obtains, after some algebra,
\begin{eqnarray}
S_p(\rho, T) &=& { 8 \pi^2 \over 3 \lambda } -
{4 \pi^2 \rho^2 \over \lambda }
\bigl\{ 2\log(\rho/2) 
\nonumber\\
& &+ 2\gamma+1 + f(T) \bigr\}
+ O(\rho^4, \rho^4/T^2), \nonumber
\end{eqnarray}
where
\[
f(T) = 8 \int { k^2 dk \over \omega_{\bf k} } { 1 \over e^{\omega_k T} - 1 }
\]
and $\gamma = 0.577..$ is the Euler constant.

It is convenient to express all relevant quantities as functions of
$T$ rather than $E$. The value of $\rho$ is determined by the
first of eqs.(\ref{stationarityS_p}),
\[
\rho^2(T) = 4 \exp \{ -2\gamma - 2 - f(T) \},
\]
while the second of eqs.(\ref{stationarityS_p}) gives
\begin{equation}
E(T) = {32 \pi^2 \rho^2 \over \lambda } \int k^2 dk
{ e^{\omega_k T} \over (e^{\omega_k T}-1)^2 }.
\label{Eperturb}
\end{equation}
Comparing this to eqs.(\ref{finalENspec}) one finds also the number of
particles,
\begin{equation}
N(T) = {32 \pi^2 \rho^2 \over \lambda } \int {k^2 dk\over \omega_{\bf k}}
{ e^{\omega_k T} \over (e^{\omega_k T}-1)^2 }.
\label{Nperturb}
\end{equation}
The exponential suppression (as a function of $T$) is determined by
\eq{Fperinst}. When the period $T$ is expressed through energy from
\eq{Eperturb}, both the suppression factor (\ref{Fperinst}) and number
of particles (\ref{Nperturb}) become functions of energy. The latter
will be shown in Fig.\ref{NUMvsPT-fig}.

\section{Numerical results}

At energies of order sphaleron energy or higher, the analytical
calculation of the multiparticle rate is not possible, and one has to
rely on numerical methods. In the case of periodic instanton, i.e.\ at
$\theta=0$, such calculations were performed previously in different
models in refs.\cite{Matveevml,HMT}. In the case $\theta \neq 0$, the
preliminary results in the scalar model (\ref{model_action}) were
published in ref.\cite{KT}.

The key step of the calculation is the numerical solution of the
boundary value problem (\ref{final_BVP}). There are several features
of this problem which make it difficult for numerical analysis. First,
\eq{BVP-fieldeq} is non-linear. Second, at $\theta\neq 0$ the field is
necessarily complex; the exponent in \eq{saddle-sigmaEN} is not
positive-definite and the solution is essentially a {\em saddle
point}.  Third, the contour ABC includes both Minkowskian (AB) and
Euclidean (BC) parts, so the numerical techniques to be used must be
suitable for both hyperbolic- and elliptic-type equations. Finally,
the initial boundary conditions (\ref{BVP-bc2}) should be imposed in
the region where the field is {\em linear}, i.e.\ the physical volume
(always finite on the lattice) must be large enough in order to allow
for linearization. The simplification to be mentioned is that
eqs.(\ref{final_BVP}) are O(3)--symmetric, so that the problem is
actually two-dimensional.

In combination, these factors impose severe constraints on both the
parameters of the lattice and numerical techniques to be used. In the
lattice version (for details see Appendix A) the boundary value
problem (\ref{final_BVP}) becomes a set of coupled non-linear
algebraic equations for the values of the field $\phi_{ij} =
\phi(t_i,r_j)$ at the lattice sites with coordinates $(t_i,r_j)$,
where $r_0,...,r_{n_x}$ correspond to the spatial radial direction,
while $t_0,...,t_{n_t}$ are complex time coordinates lying on the
contour ABC in the complex time plane. The lattice size is
characterized by three parameters: $T_M$ and $T$, the Minkowskian and
Euclidean sizes of the time contour, respectively, and the spatial
size $L$. On the Euclidean part of the contour, the solution is
compact in space. On the Minkowskian part, when viewed as evolving
backward in time from B to A, the solution propagates along the light
cone $|t|=r$. The spatial size of the lattice is therefore determined
not by the characteristic size of the Euclidean field configuration,
but by the requirement
\[
L\gsim T_M,
\]
where $T_M$ in turn is determined by the linearization
time. The latter is model--dependent\footnote{Linearization is reasonably
fast in four-dimensional models due to the volume factor. In two
dimensions it can be very slow~\cite{Rebbi-Singleton}.}. In the model
(\ref{model_action}) $T_M$ can be taken as small as $T_M = 2.5 m^{-1}$.
The corresponding value of $L$ in our calculations was $L=3 m^{-1}$.

Since the minimal physical volume is fixed in a given model, the grid 
resolution is determined by the number of mesh points in time and space 
directions, $n_t$ and $n_x$, respectively. In the $E$--$N$ space, the two 
regions of interest are large $E$ at fixed $N$ and small $N$ at fixed $E$, 
both corresponding to large values of the average momentum of initial 
particles $k_{in} \sim E/N$. The latter has to be much smaller than the 
maximum lattice momentum $\sim \pi n_x/L$, which implies 
\[
{E\over N} {L\over \pi n_x} \ll 1.
\]
Thus, $n_x$ directly controls the available region of parameter space. 

The numerical method to solve the set of equations which constitute
the lattice version of the boundary value problem (\ref{final_BVP})
should be chosen according to the specifics of the problem described
above. To get rid of the non-linearity we employ the multidimensional
analog of the Newton-Raphson method~\cite{NumRec} which approaches the
desired solution iteratively. At each iteration, the {\em linearized}
equations in the background of the current approximation have to be
solved. The next approximation is obtained by adding the solution to
the background, and the procedure is repeated. The advantage of the
algorithm is that it does not require positive-definiteness of the
matrix of second derivatives. It is, however, sensitive to zero
modes. In the absence of zero modes, the algorithm converges
quadratically; the accuracy of $10^{-9}$ is typically reached in 3-5
iterations. The convergence slows down in the presence of very soft
modes, as typically happens near bifurcation points.

The complication is that, in continuous formulation, the boundary
value problem (\ref{final_BVP}) {\em does have} an exact zero mode. 
This zero mode corresponds to translation in the real time direction
(both field equations and the boundary conditions are invariant under
such a translation). In the lattice version, due to discretization and
finite volume effects, this zero mode transforms into a very soft
one. In order to avoid the convergence problems related to this soft
mode we impose a constraint which breaks the translational symmetry
(for details see Appendix A). 

At each Newton-Raphson iteration one has to solve the set of
$n_t\times n_x$ linear equations of the general form 
\[
L\cdot u = d,
\]
where $u$ is the vector formed of $n_tn_x$ unknowns, $L$ is the matrix
of dimension $n_tn_x\times n_tn_x$ (first variation of the full
non-linear equations) and $d$ is a constant vector (full equations
evaluated at the current background; at the desired solution $d=0$).
The matrix $L$ is neither positive-definite nor even symmetric, but
has a special sparse structure as it originates from the second order
differential equation. The inversion of this matrix is the most time
consuming part of the calculation; its effectiveness determines how
large $n_t$ and $n_x$ can be taken. In our calculation we used the
forward elimination and back-substitution algorithm described in
Appendix B, which amounts to $\sim n_t n_x^3$ multiplications and
requires the storage space for $\sim n_t n_x^2$ double precision
numbers. Note that this algorithm is asymmetric in $n_t$ and $n_x$ and
is suitable for the case $n_t\gg n_x$. The results presented below
were obtained at $n_x = 80$ and $n_t=400-500$. We have been able to
implement a more efficient algorithm from the conjugate-gradient
family~\cite{NumRec} only in the case $\theta=0$.

The Newton-Raphson method requires a good initial approximation for
the solution. This favors the following general strategy. We first
find the periodic instanton solution (which corresponds to $\theta=0$)
with a given period $T$ close to the period of oscillations in the
sphaleron negative mode. It can be approximated by the sphaleron
configuration plus oscillation in the negative mode (both the
sphaleron and its negative mode have to be known for a given lattice;
the initial amplitude of the oscillation has to be chosen by trial
and error). After the periodic instanton is found, we change
parameters $T$ and $\theta$ by small steps, using the solution from
the previous run as a starting configuration for the next one. At each
step we calculate the energy $E$, number of particles $N$ and the
exponential suppression factor $F(\epsilon,\nu)$. By making use of
smoothness of $E$ and $F$ as functions of $T$ and $\theta$, the
changes in the parameters can be organized so as to maintain $E$ or
$F$, whichever is desired, (approximately) constant.

\ifinlinefigures
\begin{figure}[ht]
\epsfig{file=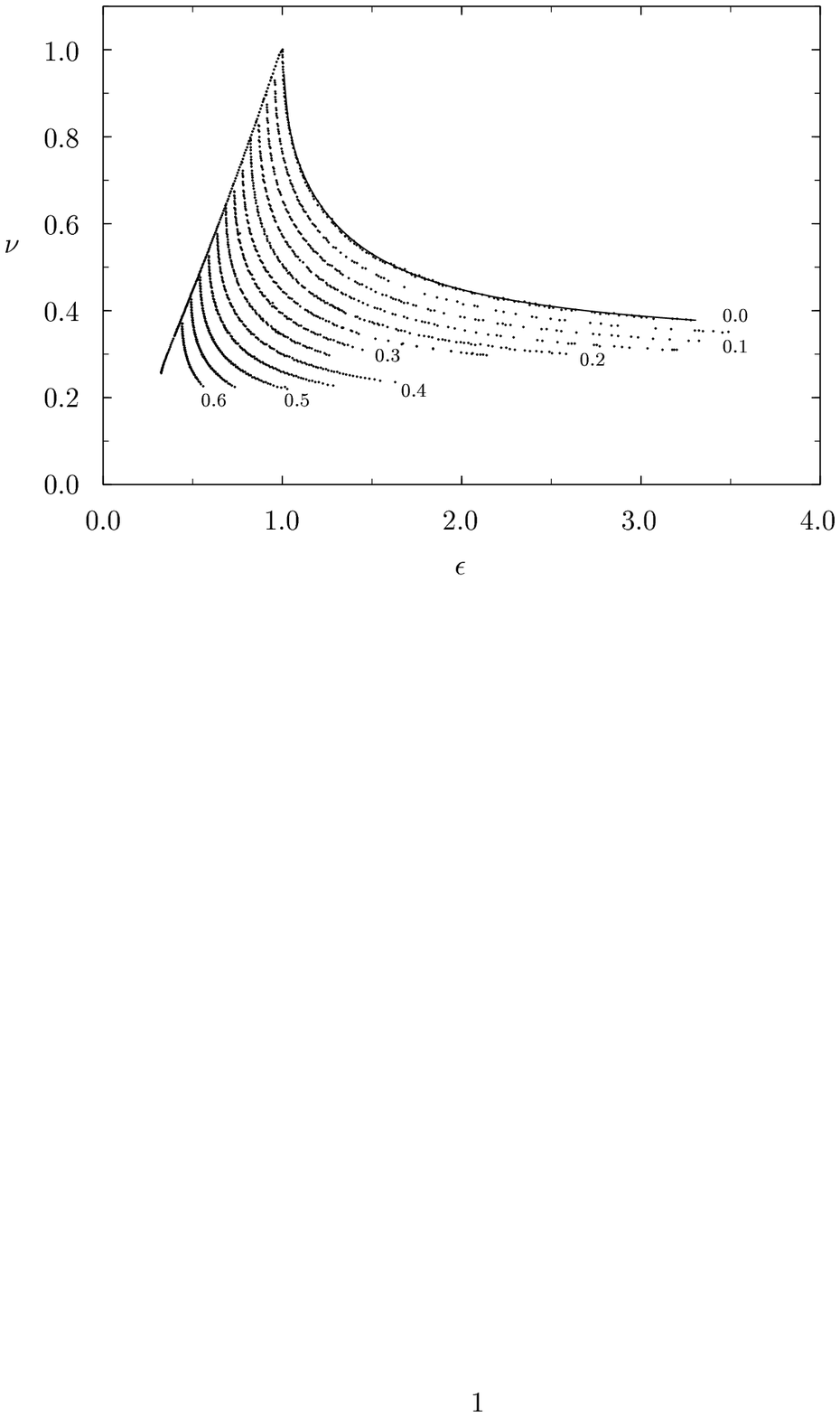,%
bbllx=92pt,bblly=490pt,%
bburx=464pt,bbury=740pt,%
width=236pt,height=159pt,%
clip=}
\caption{{\protect\small Lines of $F(\epsilon,\nu)=\const$. The value
of $F$ changes from $F=0$ for the uppermost (solid) line to $F=0.6$ with
the step $0.05$, in the normalization $F(0,0)=1$. The line directed
from the sphaleron ($\epsilon=\nu=1$) to the zero energy instanton
($\epsilon=\nu=0$) is formed by the periodic instanton solutions. Each
point on the diagram represents the solution to the boundary value
problem (\protect\ref{final_BVP}).}}
\label{FofEN-fig}
\end{figure}
\fi
The behavior of the multiparticle rate we have found in the model
(\ref{model_action}) is summarized in Fig.\ref{FofEN-fig} which shows
the lines of constant $F$ in the $\epsilon$--$\nu$ plane.  The value
of $F$ changes from $F=0$ for the uppermost (solid) line to $F=0.6$
for the lowest one, with the step 0.05. Units are such that the zero
energy instanton suppression is $F(0,0)=1$. Each dot in
Fig.\ref{FofEN-fig} represents the solution to the boundary value
problem (\ref{final_BVP}) with corresponding $\epsilon$
and $\nu$. For completeness, Fig.\ref{FofEN-fig} also shows the
periodic instanton solutions which form the line directed from the
sphaleron $\epsilon=\nu=1$ to the zero energy instanton
$\epsilon=\nu=0$.

The behavior of the function $F$ is rather remarkable. Near the
periodic instanton, the dependence of $F$ on $\nu$ is very weak.
The number of particles in the initial state can be
substantially reduced with almost no increase in the exponential
suppression. At the very periodic instanton, i.e.\ at $\theta=0$, the
slope of the lines of constant $F$ becomes infinite. The latter is
easy to see analytically from \eq{saddle-sigmaEN} by making use of the
fact that $F$ is stationary with respect to variations of $T$ and
$\theta$. One finds
\[
{dN\over dE}\Bigm|_{F=\mbox{\tiny const}} = - {T\over \theta}.
\]
The r.h.s. goes to infinity at $\theta\to 0$.

The behavior of the function $F$ changes as one moves away from the
periodic instanton.  At small $\nu$ or at high energies, the dependence
of $F$ on energy becomes weak. Unlike the vicinity of the periodic
instanton, in this region the increase of energy does not lead to a
noticeable reduction in the exponential suppression.

As follows from Fig.\ref{FofEN-fig}, the dependence of $F$ on the
number of particles at fixed energy is monotonic, in agreement with
general arguments of ref.\cite{RT,T}.
Thus, the two-particle cross section
\[
\sigma_{tot}(E) \sim \exp \left\{ -
{8\pi^2\over 3\lambda}F(\epsilon,0)\right\}
\]
is certainly exponentially suppressed at $E\lsim 3.5 E_{sph}$.  It is
clear from Fig.\ref{FofEN-fig} that the energy $E_*$ (if any) at which
the exponential suppression may disappear is substantially higher.  A
simple estimate is obtained by taking the average slope of the line
$F=0$ in the energy range $2\lsim \epsilon\lsim 3$ and performing
linear extrapolation. One finds that the extrapolated line $F=0$
crosses $\nu=0$ axis at the point $\epsilon_* \approx 10$. This gives
the following lower bound for the energy $E_*$,
\[
E_* > 10 E_{sph}.
\]
This value is likely to be underestimated.

\ifinlinefigures
\begin{figure}[ht]
\epsfig{file=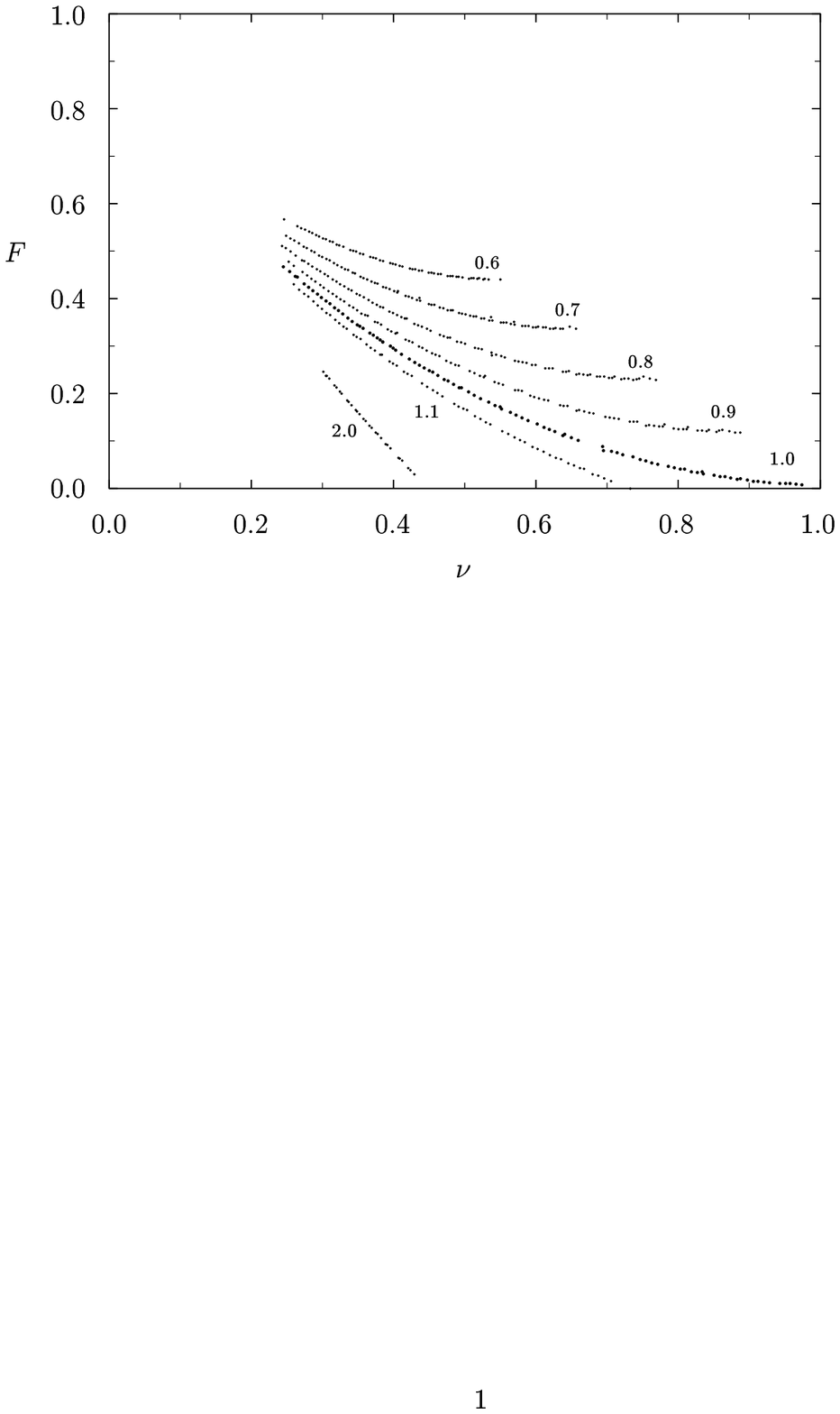,%
bbllx=87pt,bblly=490pt,%
bburx=459pt,bbury=740pt,%
width=236pt,height=161pt,%
clip=}
\caption{\protect\small $F(\epsilon,\nu)$ as a function of $\nu$ at
fixed $\epsilon$. Numbers near the curves show the values of 
$\epsilon$.}
\label{FofN-fig}
\end{figure}
\fi
Our present data allow a rough estimate of the exponential suppression
of the two-particle cross section in the region $E\sim E_{sph}$. This
estimate can be obtained by extrapolating the function $F$ to
$\nu=0$. Fig.\ref{FofN-fig} shows the dependence of the function $F$
on $\nu$ for various values of $\epsilon$ in the range $\epsilon\sim
1$. Different curves of Fig.\ref{FofN-fig} correspond to $\epsilon =
0.6, 0.7, 0.8, 0.9, 1.0,1.1, 2.0$ from top to bottom. The estimated
value is
\[
F_{HG}(\epsilon=1) \approx 0.8
\]
with the accuracy of order 10\%. Fig.\ref{FofN-fig} confirms the
conclusion that the dependence of $F_{HG}(\epsilon)$ on $\epsilon$ is
slow: all curves, including the lowest one corresponding to
$\epsilon=2$, converge as $\nu$ goes to zero and point roughly to the
same value around $0.8$. One concludes that at $E\sim E_{sph}$ the
zero energy instanton suppression is reduced by only about 20\%. For
more accurate estimate of the value of $F_{HG}(\epsilon)$ more data
are needed in the region of small $\nu$.

\section{Discussion and conclusions}

Since no analogous calculations have been performed previously, it is
worth to discuss in more detail the interpretation of the numerical
results. The first point to be considered is their correspondence to
the continuum theory. The algorithm we used in numerical calculations
is asymmetric in $n_t$ and $n_x$; it mainly restricts the space grid
resolution. So, the value of $n_x$ is our main concern.

\ifinlinefigures
\begin{figure}[ht]
\epsfig{file=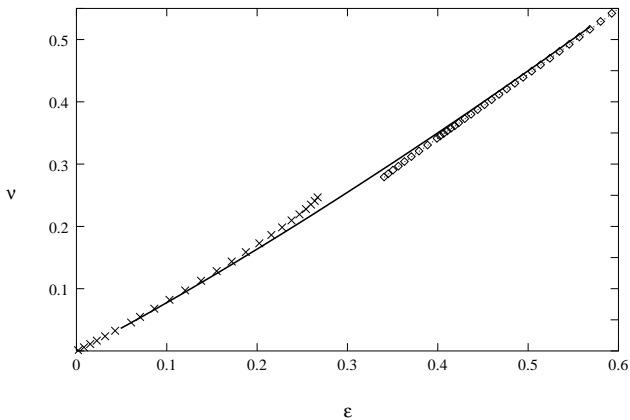,%
width=236pt,height=157pt}
\caption{\protect\small Low energy perturbation theory for the
periodic instanton ($\times$) versus numerical results at $n_x=80$
($\diamond$) and $n_x=240$ (solid line) space points.}
\label{NUMvsPT-fig}
\end{figure}
\fi
The results presented in the previous Section were obtained on the
lattice with $n_x=80$ space points in the space volume $L=3m^{-1}$. In
Fig.\ref{NUMvsPT-fig} we plot the line of periodic instantons from
Fig.\ref{FofEN-fig} (shown by diamonds) versus analytical expressions
of Sect.4 (shown by crosses). The region of validity of perturbation
theory does not overlap with $n_x=80$ data. Fortunately, in the case
$\theta=0$ the grid resolution can be increased. In
Fig.\ref{NUMvsPT-fig} the solid line represents periodic instantons
obtained numerically on the lattice with $n_x=240$ points. This line
matches both the $n_x=80$ data and perturbation theory. It is clear
from Fig.\ref{NUMvsPT-fig} that $n_x=80$ data are accurate above
$\nu=0.35$, at least in the case $\theta=0$. Fig.\ref{NUMvsPT-fig}
also indicates the region of validity of the perturbative
results, eqs.(\ref{Eperturb})-(\ref{Nperturb}).

\ifinlinefigures
\begin{figure}[ht]
\epsfig{file=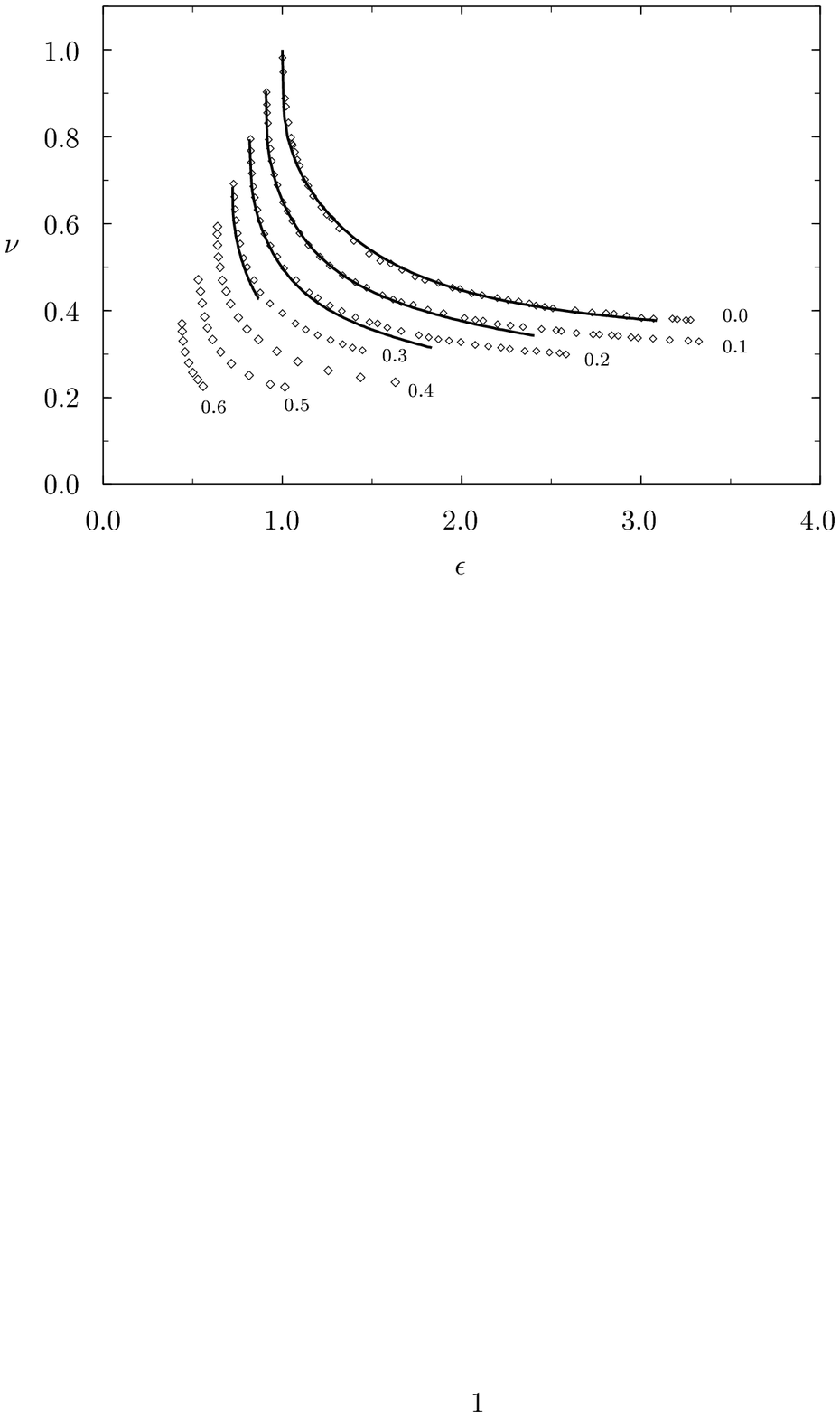,%
bbllx=87pt,bblly=480pt,%
bburx=459pt,bbury=750pt,
width=236pt,height=175pt,%
clip=}
\caption{{\protect\small Lines of constant $F$ at two grid
resolutions:  $n_x=80$ ($\diamond$) and  $n_x=40$ (solid lines). }}
\label{40vs80-fig}
\end{figure}
\fi
In the general case $\theta\neq 0$ we cannot match the perturbative
and numerical results because we do not have numerical data at
sufficiently small $\nu$ and $\epsilon$. In order to check the
dependence on grid resolution, we plot in Fig.\ref{40vs80-fig} the
$n_x=80$ data versus $n_x=40$ data of ref.\cite{KT}. The $n_x=40$ data
are shown by solid lines. The agreement is good for $\nu\gsim 0.35$
and gets worse for smaller $\nu$, in accord with our previous
estimate. The important observation is that the high-resolution curves
lie {\em above} the low-resolution ones.  Thus, both the exponential
suppression of the two-particle cross section and the energy $E_*$ at
which it may become exponentially unsuppressed are somewhat higher in
the continuum limit than follows from our lattice results.

We turn now to the interpretation of the numerical solutions presented
in Fig.\ref{FofEN-fig} as those describing the false vacuum decay.  As
mentioned in Sect.2, there may be several solutions to the boundary
value problem (\ref{final_BVP}); the physically relevant one is that
continuously connected to the low energy periodic instanton. Since
Fig.\ref{FofEN-fig} demonstrates perfectly smooth behavior and
suggests the absence of bifurcations in the scanned region of
parameters, we believe that the solutions of Fig.\ref{FofEN-fig} indeed
describe the false vacuum decay. Here we consider more direct
argument.

As explained in Sect.2, the real time part of each solution describes
the evolution of the system after tunneling. By looking at the final
field it is possible, in principle, to determine whether the false
vacuum decay took place. In the model (\ref{model_action}) the clear
signature of the decay is a singularity of the field on the positive
part of the real time axis (see Fig.\ref{phi4sings-fig}).
\ifinlinefigures
\begin{figure}[ht]
\begin{center}
\epsfig{file=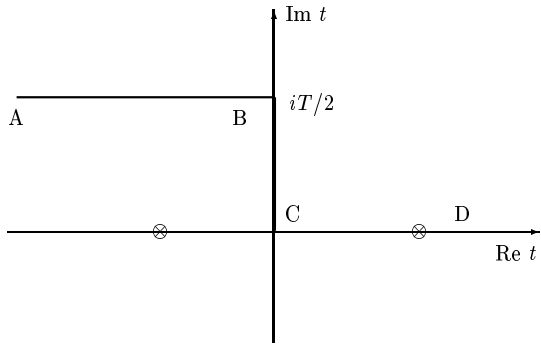,%
bbllx=185pt,bblly=625pt,%
bburx=430pt,bbury=815pt,%
width=210pt,height=160pt,%
clip=}
\end{center}
\caption{{\protect\small The position of singularities (shown by
crossed circles) of the solution to the boundary value problem
(\protect\ref{final_BVP}) in the $-\lambda\phi^4$ model. The boundary 
value problem is solved on the contour ABC. }}
\label{phi4sings-fig}
\end{figure}
\fi

Since the boundary value problem is solved on the contour ABC, the
real time part of the solution has to be found separately by solving
(numerically) the initial data problem along the real time axis, with
initial values of the field and its time derivative determined by the
solution on the contour ABC. We have performed such calculations and
found that most of the solutions of Fig.\ref{FofEN-fig} indeed have
singularity on the positive part of the real time axis, whereas some
solutions, in particular those with $\epsilon \gsim 1.5$, apparently
do not have such a singularity. There is, however, a problem with the
interpretation of the results of this check. The reason is that the
initial data problem mentioned above is unstable.
  
Let us consider the situation in more detail. As explained in Sect.2,
the solutions to the boundary value problem (\ref{final_BVP}) have
specific analytic structure which requires the presence of singularity
between the negative part of the real time axis and the contour
ABC. In the model (\ref{model_action}) this singularity lies at real
negative time\footnote{We believe that this is specific to models with
the potential unbounded from below.} as shown in
Fig.\ref{phi4sings-fig}.  Since the boundary value problem is
invariant under translations in the real time direction, only the
distance between the two singularities of Fig.\ref{phi4sings-fig} has
invariant meaning.  In numerical calculations the translational
invariance is fixed by the constraint in such a way as to keep the
left singularity at approximately the same place far from the
asymptotic region A and not too close to the origin. All solutions of
Fig.\ref{FofEN-fig} have left singularity in the range $-0.5 < t <
-0.3$.

As moving from left singularity of Fig.\ref{phi4sings-fig} to the
right one along the real time axis, the field comes from infinity,
bounces off the potential barrier and goes back. At $E < E_{sph}$ this
always takes finite amount of time. At $E > E_{sph}$ there exist
solutions which spend long time oscillating above the
sphaleron. Clearly, these solutions are unstable; small perturbations
may cause them to roll finally to the false vacuum instead of going
back to infinity. Numerical experiments show that the instability is
very strong and gets worse for higher energies. Because of this
instability, the position of the right singularity cannot be
determined reliably when the distance to the singularity is larger
than a certain amount, typically $\simeq 0.7$ for low energies and
$\simeq 0.5$ for high energies. This can be checked by studying the
dependence of the distance between the singularities on the constraint
for given $E$ and $N$. When the constraint is imposed in such a way
that the singularities are at the same distance from the origin, their
positions are determined better.  In the asymmetric case, the measured
distance between singularities is always larger and the right
singularity may even disappear.

At low energies and far from the line $F=0$, the distance between
singularities is small enough to be determined reliably. In this
region the right singularity is always present and the distance
between the singularities does not depend on the constraint. At high
energies or close to the line $F=0$, the distance is larger and
becomes constraint--dependent, so that its real value cannot be
determined by the above method. On the $(E-N)$ plane there is no
well-defined boundary between the two regions. This makes us to infer
that the interpretation of our solutions as describing the false
vacuum decay is correct.

Let us come to conclusions. Numerical calculations we have performed
demonstrate that the formalism of refs.\cite{RT,T,RSTboundary} indeed
provides a practical way to calculate the exponential suppression of
the two-particle cross section of induced tunneling,
$F_{HG}(\epsilon)$. The auxiliary object which we actually calculated,
the logarithm of the multiparticle probability $F(\epsilon,\nu)$, has
regular behavior consistent with theoretical expectations. This
allows to set lower bounds on energy $E_*$ at which the two-particle
cross section may become exponentially unsuppressed, and to estimate
the value of the suppression at energies $E\sim E_{sph}$.  We believe
that analogous calculations are possible in any model with tunneling
transitions.

In the $-\lambda\phi^4$ model we have found that the exponential
suppression of the total cross section of induced false vacuum decay
persists at least to energies of order $10 E_{sph}$. At $E\sim
E_{sph}$ about $80\%$ of the zero energy suppression is still
present. The accuracy of this estimate crucially depends on the grid
resolution, as can be seen from Fig.\ref{40vs80-fig}.

Finally, let us comment on the similarity between the $-\lambda\phi^4$
model and the SU(2) Yang-Mills-Higgs system. It arises from the softly
broken conformal symmetry present in both cases. Due to this (broken)
symmetry, the low energy tunneling in both cases is dominated by
constrained instantons whose size tends to zero in the limit $E\to
0$. The multiparticle rates should also have similar behavior, at
least in the low energy domain. Therefore, we expect that in the SU(2)
Yang-Mills-Higgs case it should be possible to calculate the function
$F_{HG}(\epsilon)$ numerically along the lines we followed in this
paper.

\acknowledgments

The authors are grateful to S.~Habib, M.~Libanov, E.~Mottola, C.~Rebbi,
V.A.~Rubakov, R.~Singleton and S.~Troitsky
for numerous discussions at different stages of this work.
The work is supported in part by Award
No. RP1-187 of the U.S. Civilian Research \& Development Foundation for
the Independent States of the Former Soviet Union (CRDF), and by
Russian Foundation for Basic Research, grant 96-02-17804a.

\appendix
\section{Lattice formulation of the boundary value problem}

In this Appendix we give the lattice formulation of the boundary value
problem (\ref{final_BVP}) and describe the method which we use to
solve it.

We first note that the boundary value problem (\ref{final_BVP}) is
O(3)-symmetric, so we may restrict ourselves to field configurations
depending on $r=\sqrt{\bm{x}^2}$ and $t$. In order to avoid the
discretization of $r$--dependent derivative terms in the action, it is
convenient to change variables according to
\[
\phi = {1\over r} \psi.
\]
The action becomes
\begin{equation}
S = 4 \pi \int dt\int\limits_0^{\infty}dr \Bigl\{
\half  {\dot\psi}^2
- \half (\partial_r \psi)^2
- \half \psi^2   + {1\over 4r^2}\psi^4 \Bigr\}.
\label{rescaled_action}
\end{equation}
In order to obtain the lattice formulation of the boundary value
problem (\ref{final_BVP}), we first derive the lattice version of the
exponent in \eq{smallPIsigmaEN} and then repeat all steps of Sect.2
leading to eqs.(\ref{final_BVP}).

Consider first the discretization of the action
(\ref{rescaled_action}). In the lattice formulation, it depends on
${(n_t+1)(n_x+1)}$ complex variables $\psi_{ij}=\psi(t_i,r_j)$, where
$r_i\in [0,L]$ with $r_0=0$, $r_{n_x}=L$, while $t_i$ are complex
numbers lying on the contour ABC of Fig.(\ref{phi4sings-fig}) so that
$t_0=-T_M+iT/2$, $t_{n_t}=0$. When $T$ is small (this is the case in
the high energy domain) the contour ABC passes close to the
singularity of the field and the solution looses the accuracy. In this
region of parameter space we deform the contour (leaving the point
$t=0$ and the initial asymptotic region intact) so as to avoid the
singularity.

Since fields are associated with the lattice sites while their
derivatives are associated with links, we define two different sets of
intervals
\[
dr_j = r_{j+1} - r_j, \hskip 0.5cm j=0\ldots n_x-1,
\]
and
\[
\tilde{dr_j} = (dr_{j-1} + dr_j)/2, \hskip 0.5cm j=1\ldots n_x-1,
\]
\[
\tilde{dr}_{0,n_x} = dr_{0,n_x-1}/2 ,
\]
and similarly for $dt_i$ and $\tilde{dt}_i$. The tilted intervals are
used to approximate integrals with the integrand defined at lattice
sites, while non-tilted ones are used for integrands defined on links.
With these definitions, the discretized action (\ref{rescaled_action})
reads
\begin{eqnarray}
S &=& 4 \pi \sum\limits_{ij} \Bigl\{
\half (\psi_{i+1,j}-\psi_{ij})^2 {\tilde{dr_j}\over dt_i}
\nonumber\\
& &- \half (\psi_{i,j+1}-\psi_{ij})^2 {\tilde{dt_i}\over dr_j}
- V_{ij}\tilde{dt_i}\tilde{dr_j}\Bigr\},
\label{discrete-action}
\end{eqnarray}
where
\begin{eqnarray}
V_{ij} &=& \half \psi_{ij}^2 -  {1\over 4r_j^2}\psi_{ij}^4 \;\;\; \mbox{at}
\;\;\; j\neq 0,
\nonumber\\
V_{i0}&=&0\, .\nonumber
\end{eqnarray}
This discretization scheme has the advantage of being symmetric and
additive: the action for the whole lattice is the sum of the actions
of elementary plaquettes. It has the accuracy $O(dr^2)$. The
boundaries are treated to the same accuracy, which is important since
the derivation of the boundary problem (\ref{final_BVP}) requires
integrations by parts.

Consider now the boundary term in \eq{smallPIsigmaEN}. On the lattice,
the plane waves are no longer exact eigenfunctions of the
Hamiltonian. To find their analog consider the quadratic part of
\eq{discrete-action} in the limit of continuous time,
\[
S^{(2)} = 4 \pi \int dt \sum\limits_j \Bigl\{
\half {\dot\psi_j}^2 \tilde{dr_j}
- { (\psi_{j+1}-\psi_{j})^2 \over 2 dr_j}
- \half \psi_{j}^2 \tilde{dr_j} \Bigr\},
\]
where $\psi_j = \psi(t,r_j)$. The kinetic term of this action takes
the canonical form in terms of variables $\chi_j(t) =
\psi_j(t)\sqrt{\tilde{dr_j}}$. The rest of the integrand can be
written as minus free lattice Hamiltonian,
\[
-\half \sum\limits_{jk} h_{jk} \chi_j\chi_k,
\]
where
\begin{eqnarray}
h_{jk} &=& \delta_{jk} \left( {1\over \tilde{dr_j} dr_{j-1}} + {1\over
\tilde{dr_j} dr_j} + 1 \right)
\nonumber\\
&&-{\delta_{j+1,k} \over dr_j\sqrt{\tilde{dr_j}\tilde{dr_k}}}
-{\delta_{j-1,k} \over dr_k\sqrt{\tilde{dr_j}\tilde{dr_k}}}.
\nonumber
\end{eqnarray}
The diagonalization of $h_{jk}$ determines the eigenfunctions
$\xi^{(n)}_k$ and eigenvalues $\omega_n^2$ which substitute the plane
waves and frequencies $\omega_{\bf k}^2$. We perform the
diagonalization numerically. In fact, all expressions of Sect.2
involving momentum representation can be readily translated to the
lattice language by means of the substitutions
\begin{eqnarray*}
\int d{\bm{k}} &\to& \sum\limits_n \\
\omega_{\bf k} &\to& \omega_n \\
\phi(t_i,\bm{k}) &\to& \sum\limits_j \xi^{(n)}_j\sqrt{\tilde{dr_j}}
\psi_{ij}.
\end{eqnarray*}
In particular, the boundary term in \eq{saddle-sigmaEN} becomes
\begin{equation}
\sum\limits_{jk}  \Omega_{jk} \left\{
-{1-\gamma \over 1+\gamma}
\Re\psi_{0j} \Re\psi_{0k}
+{1+\gamma \over 1-\gamma}
\Im\psi_{0j} \Im\psi_{0k}
\right\}
\label{lat-BT}
\end{equation}
where
\[
\Omega_{jk} = \sum\limits_n\sqrt{\tilde{dr_j}} \xi^{(n)}_j\omega_n
\xi^{(n)}_k\sqrt{\tilde{dr_k}} .
\]
The matrix $\Omega_{jk}$ is calculated numerically. Note that it has
to be calculated only once for a given spatial lattice $\{r_j\}$.

Now we are in position to derive the lattice version of the boundary
value problem (\ref{final_BVP}). The lattice analog of the field equation
$\delta S/\delta\phi=0$ is
\begin{equation}
\dpar{S}{\psi_{ij}}=0
\label{lat-fieldeq}
\end{equation}
where $i=1...n_t-1$. The final boundary conditions immediately
translate to
\begin{eqnarray}
\Im \psi_{n_t,j}&=&0,
\nonumber\\
\Im \dpar{S}{\psi_{n_t,j}} &=& 0.
\label{lat-bc1}
\end{eqnarray}
In order to derive the initial boundary conditions,
one has to consider the lattice version of the exponent in
\eq{smallPIsigmaEN}, take the derivative with respect to $\psi_{0j}$
and then set $\psi_{0j}'=(\psi_{0j})^*$ and $\gamma_{\bf k} =\gamma = 
\e^{-\theta}$ as discussed in Sect.2. The result reads
\begin{equation}
\dpar{S}{\psi_{0j}} +
\sum\limits_k \Omega_{jk}\left\{ i{1-\gamma\over 1+\gamma}\Re \psi_{0k} -
{1+\gamma\over 1-\gamma}\Im \psi_{0k} \right\} = 0
\label{lat-bc2}
\end{equation}
Note that total number of equations matches the number of unknowns.
Eqs.(\ref{lat-fieldeq})-(\ref{lat-bc2}) form a set of coupled
non-linear algebraic equations which constitute the lattice analog of
the boundary value problem (\ref{final_BVP}).

As has been noted in Sect.5, in continuum version these equations are
invariant under translation in the real time direction, which leads to
the existence of almost zero mode on the lattice, i.e.\ continuous family of
(approximate) solutions to eqs.(\ref{lat-fieldeq})-(\ref{lat-bc2}). 
This spoils the convergence in the Newton-Raphson method and has to be
cured. The standard trick is to introduce the constraint which breaks
the translational invariance. In the vicinity of the sphaleron it is
natural to require that at the point $t=0$ the field velocity has zero
projection on the sphaleron negative mode, 
\begin{equation}
\sum\limits_j  
\label{constraint}\Xi_j^{(-)} \sqrt{\tilde{dr}_j}
\dpar{S}{\psi_{n_t,j}} =0 ,
\end{equation}
where $\Xi_j^{(-)}$ is the sphaleron negative mode on the lattice.
Far from the sphaleron there is no natural choice, so we use the
constraint (\ref{constraint}) at all values of parameters. In the
continuum theory physical quantities do not depend on the constraint.

\section{Numerical algorithm}

Here we give basic formulae for the forward elimination -- backward
substitution algorithm which we use to solve the linear equations
arising at each Newton-Raphson iteration. 

The general form of the linearized equations is 
\begin{equation}
L \cdot u = d,
\label{Lu=R}
\end{equation}
where $L$ is the matrix of first derivatives of the full equations,
$u_{ij}$ are unknowns, while $-d_{ij}$ are full equations calculated
at the current background ($d_{ij}=0$ if the background is a
solution). The matrix $L$ has the dimension $(n_t+1)(n_x+1)\times
(n_t+1)(n_x+1)$. It is a sparse matrix with a special structure which
is most conveniently represented in the following form. Let us
introduce vector notations in which the spatial index $j$ is implicit,
$\bm{u}_i = \{ u_{i0}, \ldots ,u_{in_x} \}$ and similarly for
$\bm{d}_i$. Then the matrix $L$ has block-three-diagonal form with the
diagonal blocks $\bm{D}_i$ and off-diagonal blocks $\bm{D}^{(-)}_i$
and $\bm{D}^{(+)}_i$, all of the same dimension $(n_x+1)\times
(n_x+1)$. Eq.(\ref{Lu=R}) is equivalent to the following set of
equations,
\begin{equation}
\begin{array}{lllllll}
& & \bm{D}_0 \bm{u}_0 \relax & + &  \bm{D}^{(+)}_0 \bm{u}_1 &=& \bm{d}_0, \\
& & & & &\ldots & \\
\bm{D}^{(-)}_i \bm{u}_{i-1} &+& \bm{D}_i \bm{u}_i &+& 
\bm{D}^{(+)}_i \bm{u}_{i+1} &=& \bm{d}_i, \\
& & & & &\ldots & \\
\bm{D}^{(-)}_{n_t} \bm{u}_{n_t-1} &+&  \bm{D}_{n_t} \bm{u}_{n_t}
& &
&=& \bm{d}_{n_t}. 
\end{array}
\label{Du+Du+Du=R}
\end{equation}
The first and last of these equations represent the boundary
conditions. 

Let us define a set of matrices $\bm{A}_i$ of dimension  $(n_x+1)\times
(n_x+1)$ and vectors $\bm{b}_i$, $i=0\ldots n_t-1$, by the equations
\begin{equation}
\bm{u}_i = \bm{A}_i \bm{u}_{i+1} + \bm{b}_i.
\label{u=Au+b}
\end{equation}
The first of eqs.(\ref{Du+Du+Du=R}) implies 
\[
\begin{array}{rcl}
\bm{A}_0 &=& - [\bm{D}_0]^{-1} \bm{D}^{(+)}_0, \\
\bm{b}_0 &=& [\bm{D}_0]^{-1} \bm{d}_0.
\end{array}
\]
Eqs.(\ref{Du+Du+Du=R}) and (\ref{u=Au+b}) together give the recursion
relation for $\bm{A}_i$ and $\bm{b}_i$, 
\[
\begin{array}{rcl}
\bm{A}_i &=& - [\bm{D}^{(-)}_i \bm{A}_{i-1} + \bm{D}_i]^{-1} \bm{D}^{(+)}_i, \\
\bm{b}_i &=& [\bm{D}^{(-)}_i \bm{A}_{i-1} + \bm{D}_i]^{-1} 
[\bm{d}_i - \bm{D}^{(-)}_i\bm{b}_{i-1}] .
\end{array}
\]
At the stage of forward elimination, all $\bm{A}_i$ and $\bm{b}_i$ are
calculated and stored (this procedure is equivalent to the elimination
of the lower block-subdiagonal of the matrix $L$).

At the last point $i=n_t$, the third of eqs.(\ref{Du+Du+Du=R}) and
\eq{u=Au+b} taken at $i=n_t-1$  determine the vector of
unknowns $\bm{u}_{n_t}$,
\[
\bm{u}_{n_t} = [\bm{D}^{(-)}_{n_t} \bm{A}_{n_t-1} + \bm{D}_{n_t}]^{-1} 
[\bm{d}_{n_t} - \bm{D}^{(-)}_{n_t}\bm{b}_{n_t-1}] .
\]
Other unknowns are found by sequential use of \eq{u=Au+b}
(back-substitution). 

Clearly, the most time consuming stage is the calculation of the matrices
$\bm{A}_i$. It amounts to matrix inversion and subsequent
multiplication of the diagonal matrix by the result, all repeated
$n_t$ times. This can be done in $\sim n_t n_x^3$ complex 
multiplications~\cite{NumRec} with a coefficient close to 1. The
algorithm requires storage space for $n_tn_x^2$ complex numbers.

\ifinlinefigures\relax
\fi


\begin{references}
\bibitem{Bounce} 
     S.~Coleman, Phys. Rev. D {\bf 15}, 2929 (1977).
\bibitem{GaugeInstantons} 
     A.~A.~Belavin, A.~M.~Polyakov, A.~S.~Schwartz and
     Yu.~S.~Ty\-u\-p\-kin, Phys. Lett. {\bf 59B}, 85 (1975).
\bibitem{KLINKHAMER-MANTON} 
     N.~S.~Manton, Phys. Rev. D {\bf 28}, 2019 (1983);
     F.~R.~Kli\-nk\-ha\-mer and N.~S.~Manton, {\it ibid.} {\bf 30}, 2212 (1984).
\bibitem{VOLOSHIN-KOBZAREV-OKUN} 
     M.~B.~Voloshin, I.~Yu.~Kobzarev and L.~B.~Okun,
     Sov. J. Nucl. Phys. {\bf 20}, 644 (1975).
\bibitem{RE}
     A.~Ringwald, Nucl. Phys. {\bf B330}, 1 (1990);
     O.~Espinosa, {\it ibid.} {\bf B334}, 310 (1990).
\bibitem{t'Hooft} 
     G.'t~Hooft, Phys. Rev. D {\bf 14}, 3432 (1976).
\bibitem{RubShap} 
     V.~A.~Rubakov and M.~E.~Shaposhnikov, Usp. Fiz. Nauk {\bf 39}, 461 (1996).
\bibitem{McLVV} 
     L.~McLerran, A.~Vainshtein and M.~B.~Voloshin, Phys. Rev. D {\bf 42}, 171 (1990).
\bibitem{KRT2}
     S.~Yu.~Khlebnikov, V.~A.~Rubakov and P.~G.~Tinyakov, Nucl. Phys.
     {\bf B350}, 441 (1991).
\bibitem{Yaffe}
     L.~Yaffe, {\it Scattering amplitudes in instanton background\/},
     In: M.~Mattis and E.~Mottola, editors, Proceedings of the Santa Fe
	     Workshop (World Scientific, Singapore, 1990).
\bibitem{ArnoldMattis} 			
     P.~B.~Arnold and M.~P.~Mattis, Mod. Phys. Lett. A {\bf 6}, 2059 (1991).
\bibitem{reviews}
     M.~Mattis, Phys. Rep. {\bf 214}, 159 (1992);
     P.~G.~Tinyakov, Int. J. Mod. Phys. A {\bf 8}, 1823 (1993).
\bibitem{RT}
     V.~A.~Rubakov and P.~G.~Tinyakov, Phys. Lett. B {\bf 279}, 165 (1992).
\bibitem{T}
     P.~G.~Tinyakov, Phys. Lett. B {\bf 284}, 410 (1992).
\bibitem{RSTboundary}
     V.~A.~Rubakov, D.~T.~Son and P.~G.~Tinyakov,
     Phys. Lett. B {\bf 287}, 342 (1992).
\bibitem{SRLiuville}
     V.~A.~Rubakov and D.~T.~Son, Nucl. Phys. {\bf B422}, 195 (1994);
     {\it ibid.} {\bf B424}, 55 (1994).
\bibitem{vac-decay}
     I.~K.~Affleck and  F.~De~Luccia, Phys. Rev. D {\bf 20}, 3168 (1979);
     M.~B.~Voloshin and K.~G.~Selivanov, Yad. Fiz. {\bf 44}, 1336 (1986); 
     M.~B.~Voloshin, Nucl. Phys. {\bf B363}, 425 (1991);
     J.~Ellis, A.~Linde and M.~Sher, Phys. Lett. B {\bf 252}, 203 (1990);
     V.~A.~Rubakov, D.~T.~Son and  P.~G.~Tinyakov, Phys. Lett. B {\bf 278}, 279 (1992). 
\bibitem{unitarity} 
     V.~I.~Zakharov, Nucl. Phys. {\bf B353}, 683 (1991);
     {\it ibid.} {\bf B383}, 218 (1992).
\bibitem{KT}
     A.~N.~Kuznetsov and P.~G.~Tinyakov, Mod. Phys. Lett. A {\bf 11}, 479 (1996).
\bibitem{Mueller} 
     A.~H.~Mueller, Nucl. Phys. {\bf B401}, 93 (1993).
\bibitem{FadSlav} 
     L.~D.~Faddeev and A.~A.~Slavnov,
     {\it Introduction to Quantum Theory of Gauge Fields}
	(Nauka, Moscow, 1978).
\bibitem{Rebbi-Singleton}
     C.~Rebbi and R.~Singleton, Phys. Rev. D {\bf 54}, 1020 (1996).
\bibitem{KRTperiod}
     S.~Yu.~Khlebnikov, V.~A.~Rubakov and P.~G.~Tinyakov, Nucl. Phys.
     {\bf B367}, 334 (1991).
\bibitem{Fubini}
     S.~Fubini, Nuovo Cimento {\bf 34A}, 521 (1976).
\bibitem{Lipatov} 
     L.~N.~Lipatov, Zh. Eksp. Teor. Fiz. {\bf 72}, 411 (1977).
\bibitem{Affleck}
     I.~Affleck, Nucl. Phys. {\bf B191}, 455 (1981).
\bibitem{NumRec} 
     W.~H.~Press {\it et al.},
     {\it Numerical Recipes in C} (Cambridge University Press,
     England, 1992);
     R.~Barrett {\it et al.}, {\it Templates for the Solution of
     Linear Systems: Building Blocks for Iterative Methods\/},
     (SIAM, Philadelphia, PA, 1994);
     C.~C.~Paige and M.~A.~Saunders, SIAM J. Numer. Anal. {\bf 12}, 617 (1975);
     ACM Trans. Math. Soft. {\bf 8}, 43 (1982).
\bibitem{sphaleron-decay} 
     M.~Hellmund and J.~Kripfganz, Nucl. Phys. {\bf B373}, 749 (1992).
\bibitem{Matveevml}
     V.~V.~Matveev, Phys. Lett. B {\bf 304}, 291 (1993).
\bibitem{HMT}
     S.~Habib, E.~Mottola and P.~G.~Tinyakov, Phys. Rev. D {\bf 54}, 7774 (1996).
\end{references}
\end{document}